\documentclass[twocolumn]{aastex6}
\usepackage{natbib}
\usepackage{hyperref}
\usepackage{colortbl}
\usepackage{color}
\newcommand{\msun}{\,{\rm M_\odot}}

\newcommand{\beq}{\begin{equation}}
\newcommand{\eeq}{\end{equation}}
\newcommand{\ba}{\begin{eqnarray}}
\newcommand{\ea}{\end{eqnarray}}
\def\spose#1{\hbox to 0pt{#1\hss}}
\newcommand{\lta}{\mathrel{\spose{\lower 3pt\hbox{$\mathchar'218$}}
      \raise 2.0pt\hbox{$\mathchar"13C$}}}
\newcommand{\gta}{\mathrel{\spose{\lower 3pt\hbox{$\mathchar"218$}}
      \raise 2.0pt\hbox{$\mathchar"13E$}}}

\definecolor{grey}{rgb}{0.75,0.75,0.75}
\definecolor{Orange}{rgb}{1.0,0.5,0.15}
\definecolor{brown}{rgb}{0.7,0.25,0.0}
\definecolor{pink}{rgb}{1.0,0.5,0.5}
\definecolor{darkerred}{rgb}{0.8,0,0}
\definecolor{darkerblue}{rgb}{0,0,0.8}
\definecolor{Blue}{rgb}{0,0.08,0.65}
\definecolor{Red}{rgb}{0.65,0.08,0.05}
\definecolor{Green}{rgb}{0.15,0.45,0.25}

\shorttitle{High-z galaxies and AGN}
\shortauthors{Volonteri et al.}

\begin{document}

\title{High-redshift galaxies and black holes in the eyes of JWST:  a population synthesis model from infrared to X-rays}

\author{Marta Volonteri\altaffilmark{1}}
\email{martav@iap.fr}
\author{Amy E. Reines\altaffilmark{2,3}}
\author{Hakim Atek\altaffilmark{1}}
\author{Daniel P. Stark\altaffilmark{4}}
\author{Maxime Trebitsch\altaffilmark{1}}
\altaffiltext{1}{Institut d'Astrophysique de Paris, Sorbonne Universit\`{e}s, UPMC Univ Paris 6 et CNRS, UMR 7095, 98 bis bd Arago, 75014 Paris, France}
\altaffiltext{2}{National Optical Astronomy Observatory, 950 North Cherry Avenue, Tucson, AZ 85719, USA}
\altaffiltext{3}{Department of Physics, Montana State University, Bozeman, MT 59717, USA}
\altaffiltext{4}{Steward Observatory, University of Arizona, 933 N Cherry Ave, Tucson, AZ 85721, USA}

\begin{abstract}
The first billion years of the Universe is a pivotal time: stars, black holes (BHs) and galaxies form and assemble, sowing the seeds of  galaxies as we know them today. Detecting, identifying and understanding the first galaxies and BHs is one of the current observational and theoretical challenges in galaxy formation.  In this paper we present  a population synthesis model aimed at  galaxies, BHs and Active Galactic Nuclei (AGNs)  at high redshift. The model builds a population based on empirical relations. The spectral energy distribution of galaxies is determined by age and metallicity, and that of AGNs by BH mass and accretion rate. We validate the model against observations, and predict properties of galaxies and AGN in other wavelength and/or luminosity ranges, estimating the contamination of stellar populations  (normal stars and high-mass X-ray binaries) for AGN searches from the infrared to X-rays, and vice-versa for galaxy searches. For high-redshift galaxies, with stellar ages $<1$ Gyr, we find that disentangling stellar and AGN emission is challenging at restframe UV/optical wavelengths, while high-mass X-ray binaries become more important sources of confusion in X-rays. We propose a color-color selection in {\it JWST} bands to separate AGN vs star-dominated galaxies in photometric observations.  We also estimate the AGN contribution, with respect to massive, hot, metal-poor  stars, at driving high ionization lines, such as \ion{C}{4} and \ion{He}{2}. Finally, we test the influence of the minimum BH mass and occupation fraction of BHs in low mass galaxies on the restframe UV/near-IR and X-ray AGN luminosity function.
\end{abstract}


\section{Introduction}
The first galaxies and black holes (BHs), within the first billion years of the Universe, have set the stage for the ensuing evolution of galaxies. Their radiation has shaped the thermal evolution of the intergalactic medium, ionizing the neutral plasma left over after electrons and protons combined to form neutral hydrogen atoms, and making the Universe transparent to UV radiation. The radiative and kinetic feedback exerted by stellar populations and supernovae, as well as by Active Galactic Nuclei (AGNs) powered by the first BHs has instead shaped the interstellar medium, influencing how stars and BHs evolve in turn, in a sometimes virtuous sometime vicious cycle, as feedback can both foster or hinder star formation and BH accretion.

Observational evidence on the first galaxies is growing \citep[for a review see][]{2016ARA&A..54..761S}, and there will be a leap forward when the {\it James Webb Space Telescope} ({\it JWST}) is launched, in the imminent future. Bright quasars have also been detected at similar cosmic epoch, when the Universe was less than a billion years \citep{Fanetal2001a,Fan2006,Venemans2013,2016ApJ...828...26M,2016ApJ...833..222J,2016ApJS..227...11B}, while the population of fainter quasars is still small \citep{Willot2010,2016ApJ...828...26M}. Currently, high redshift galaxies and quasars are studied almost separately. We know, however, that in the local Universe between a completely stellar dominated galaxy and a quasar all sorts of shades are possible. Faint AGNs are now identified even in many dwarf galaxies \citep[for a review see][]{2016PASA...33...54R}.

At high redshift, $z\gtrsim6$, the presence or absence of AGNs in the bulk of galaxies is a subject of debate, with few convincing candidates to date \citep{2013ApJ...778..130T,2015MNRAS.448.3167W,2015A&A...578A..83G,2016ApJ...823...95C,2016MNRAS.463..348V}, with searches focused mostly in the X-rays. The small number of bona-fide AGN in Lyman break galaxies is, however, consistent with theoretical expectations when realistic assumptions are made: \cite{2016ApJ...820L...6V} predict that BHs might just be smaller/fainter and below the detection limits when adopting BH-stellar mass relations appropriate for low-mass galaxies \citep{2015ApJ...813...82R}. 

The recent discovery of high ionization lines in UV-selected galaxies is opening a new way of searching for AGNs, or at least of interpreting the relative role of AGNs and hot stars as the powering mechanism. \cite{2015MNRAS.454.1393S} revealed detection of \ion{C}{4} in one of eleven known Lyman-$\alpha$  emitters at $z>7$, while \cite{2015ApJ...808..139S} and \cite{2016arXiv160900727B} discuss the detection of \ion{He}{2} and OIII] in a bright Lyman-$\alpha$ emitter at $z=6.6$. \cite{2017ApJ...836L..14M} revealed \ion{C}{4} in a gravitationally lensed Lyman-alpha emitter at $z\sim6$.  In particular, in \cite{2017ApJ...836L..14M}, they showed how UV line ratios could be used to distinguish between AGNs and massive, hot, metal-poor  stars as a powering mechanism using dedicated emission line models \citep{2016MNRAS.456.3354F}.  \cite{2017ApJ...836L..14M} argued that based on the presence of \ion{C}{4} and lack of \ion{He}{2}, this source was likely to have a break in the ionizing spectrum between 47.9 and 54.4 eV, consistent with a stellar ionizing spectrum and inconsistent with an AGN power law spectrum.   The fact that they detect strong OIII] provided further evidence in favor of metal poor hot stars, as an AGN spectrum would likely have weaker OIII] given that oxygen is triply ionized.   

In this paper we follow on the issue by estimating how often a high-z galaxy hosts a BH, an active BH, and what is the relationship between the two, not only in terms of physical properties, but also in terms of observability. \cite{2015MNRAS.454.3771P}, \cite{2016MNRAS.459.1432P} and \cite{2016arXiv161005312N} recently analyzed the observational properties of ``seed" BHs in primeval galaxies; we here take a broader view, moving to later cosmic times and including more mature galaxies and BHs. \cite{2011ApJ...733...31H} and \cite{2012ApJ...760...74H} can be seen as observational counterparts to this paper: rather than {\it assuming} that galaxies are unimportant for quasars, and AGNs are unimportant for galaxies, we {\it assess} their relationship within a statistically relevant population.  

In this paper, the first of a series, we present our methodology and a survey of the main results at $z=6$. We first calibrate our model, and provide an interpretation to existing observations in its light. We then advance predictions for galaxies and AGNs in the eyes of {\it JWST}, and finally compare the power of optical/near-IR observations versus X-rays in recovering the population properties. 

\section{Method}
We create a population of galaxies, BHs and AGNs starting from the galaxy stellar mass function. Each galaxy is assigned a metallicity and a star formation rate based on empirical relations with stellar mass  (see section 2.2).  Black holes are assigned to galaxies by assuming a relation between black hole and stellar mass, and a luminosity to an active black hole by assuming a duty cycle and a distribution of accretion rates for black holes. The duty cycle is defined as the fraction of black holes with Eddington ratio above 1\%, and the distribution of accretion rates pertains only to these black holes, i.e., those with Eddington ratio above 1\%. The model is constrained by matching the AGN and galaxy luminosity function  (see section 2.1). The model is presented in this paper at z=6 only. It is not conceived as an evolutionary, but an empirical model; this means that the free parameters need to be fit independently for each redshift where sufficient information for fitting the free parameters is available. 

We follow here the approach of \cite{2016ApJ...820L...6V}, based on  \cite{2011A&A...535A..87S}. We denote BH masses as $\mu=\log M_{\rm BH}$,   stellar masses $s$, with $s=\log M_*$, and the AGN luminosity as $l=\log L_{\rm AGN}$. We adopt a simple  functional form for the scaling between BH mass and galaxy stellar mass, $\mu= \gamma + \alpha s$, with log-normal intrinsic scatter $\sigma_{\mu}$.

Specifically, we adopt the relationship found by \cite{2015ApJ...813...82R} for moderate-luminosity AGNs, typically in lower-mass host galaxies:
\begin{equation}
\mu=(1.05\pm0.11)(s-11)+(7.45\pm 0.08),
\label{fitRV}
\end{equation}
which was shown to produce BH populations in agreement with constraints by \cite{2016ApJ...820L...6V}. $\sigma_{\mu}$ was measured in \cite{2015ApJ...813...82R} to be 0.55 dex, but here we leave $\sigma_{\mu}$ as a free parameter to be set by fitting the AGN luminosity function (LF). We do not include any redshift evolution, as whether there is an evolution, and whether it is in normalization or slope is still uncertain, both theoretically \citep{2011MNRAS.417.2085V,2015MNRAS.454..913D,2016MNRAS.460.2979V,2017arXiv170107838B} and observationally \citep[][and references therein]{2017arXiv170302041D}. 

For a given galaxy mass we assign a BH mass based on Eq.~\ref{fitRV}, and a luminosity through the probability distribution of the logarithmic Eddington ratio $\lambda$, recalling that $l=38.11+\lambda+\mu$. We consider here a lognormal distribution, motivated by observational \citep{2009MNRAS.397..135K,2012MNRAS.425..623L} and theoretical arguments \citep{2016MNRAS.460.2979V}. We refer the reader to \cite{2016arXiv160300823T} for an exhaustive discussion, and \cite{2016ApJ...826...12J} for a different perspective. We set the two parameters $\bar{\lambda}$, $\sigma_{\lambda}$ by fitting the AGN LF. Finally, we consider a duty cycle, ${\cal D}$, giving the fraction of BHs that are active. In this case we define ``active" as the fraction of BHs that are accreting at $\lambda\geqslant-2$.  In this paper, where we are mainly interested in AGNs around and below the knee of the LF, we set ${\cal D}=0.25$, based on simulation ``D" by \cite{2016arXiv160509394H}, which best reproduces observational constraints.

Our starting point is the galaxy mass function (MF)  and we create a Monte Carlo simulation of the galaxy+AGN population by assigning a BH mass ($\mu$) and luminosity ($\l$) through the Eddington ratio ($\lambda$). We also assume that only a fraction ${\cal D}$ of the BHs are in a luminous phase. With this approach we can build LFs and samples matching mass/luminosity cuts. 
 
A fraction of AGNs are obscured, and they are missed by observations. When necessary, we include a luminosity dependent correction\footnote{Recent results, based on a local {\it Swift}/BAT sample, suggest that obscuration fraction is related to Eddington ratio, rather then luminosity \citep{2017Natur.549..488R}. As the authors acknowledge therein, higher gas fractions and a more turbulent medium in high-z galaxies may induce deviations from this local result. } for  the obscured fraction based on \cite{2014ApJ...786..104U}.  The obscured fraction appears to increase with redshift up to, at least, $z\sim2-3$ \citep[e.g.,][]{2014MNRAS.437.3550M,2014MNRAS.441.1059V,2015ApJ...802...89B}. The evolution at higher redshift  is less constrained, and appears, perhaps, to saturate at $z\sim4$  \citep{2017arXiv170300657L}, but it may be affected by incompleteness at faint luminosities (Vito et al. in prep), making a robust assessment hard. For simplicity we adopt the obscured fraction at $z\sim2$ also at higher redshift, $z=6$, but, based on the studies above, it may underestimate the obscured fraction. The correction we adopt is based on an X-ray sample, and we use it also in optical/UV. \cite{2014MNRAS.437.3550M} show that in about 70\% of the sources they study the optical and X-ray classification of obscured/unobscured agrees. For the remaining 30\%, at low luminosities X-ray unobscured AGNs are obscured in optical/UV, while at high-luminosities optical/UV unobscured AGNs may have absorbed X-ray spectra.  At low-luminosities the optical/UV obscured characterization is likely induced by line-emission dilution into the dominant host-light, while at high-luminosities the X-ray obscuration may be induced by a higher gas component inducing the obscuration (which does not affect the optical range). When comparing to the X-ray LF, we do not correct for Compton thin AGNs,  with column density $N_{\rm H}=10^{22}-10^{24} \rm{cm}^{2}$, as at $z\gtrsim6$ obscuration should be negligible,  although obscuration at the level of $N_{\rm H}=10^{23}-10^{24} \rm{cm}^{2}$ cannot be ruled out, but we correct for Compton thick sources, with  column density $N_{\rm H}>10^{24} \rm{cm}^{2}$, which would still be missed. Such a population is expected to account for 30-50\% of the AGN population, based on lower redshift hard X-ray observations \citep[][and references therein]{2014ApJ...786..104U} and synthesis of the X-ray background \citep{2007A&A...463...79G}. Specifically, we follow here \cite{2014ApJ...786..104U} for self-consistency with the absorbed fraction.

In most of this work, we assign a BH in each and every galaxy, although BHs are not necessarily expected to be ubiquitous in low-mass galaxies, and we have adopted a minimum BH mass of $10^2 \msun$. These assumptions should be treated as free parameters, and in section~\ref{sec:seed} we discuss their importance and variations on the basic models. 

Several different measurements and analytical fits to the galaxy stellar MF can be found in the literature. Many of them are summarized in \cite{2013ApJ...770...57B} and  \cite{2014ARA&A..52..415M}, and \cite{2016ARA&A..54..761S} for a focus on high redshift, $z>4$, where differences and uncertainties are discussed (see Fig.11 in Madau \& Dickinson 2014).  In the following we  use as a reference the MF by \cite{2016ApJ...825....5S}, extrapolated down to $10^5 \msun$ and up to $3\times10^{12} \msun$.  The choice of the low-mass end does not impact the results,   in the sense that the luminosity functions are not strongly dependent on the minimum value of stellar mass, as long as there is no change in the slope of the MF at the faint end. Such slope change has not been yet reported in the literature (but the LF is eventually expected to bend). The bright end of the AGN LF is sensitive to the maximum galaxy mass we consider. In practice, we have to include galaxies up to, at least,  the stellar mass at which the stellar mass function, multiplied by the duty cycle, drops to levels consistent with the bright end of the AGN mass function. In this paper we focus on a specific redshift, $z=6$, as this is the highest redshift where we have statistical information on the AGN LF. The galaxy LF/MF are currently measured out to $z=8$, and the model can be extrapolated to the same redshift in order to make predictions for future observations. 

\begin{figure}[!h]
\hspace{-.4cm}
\vspace{-1.4cm}
\includegraphics[width=3.5in]{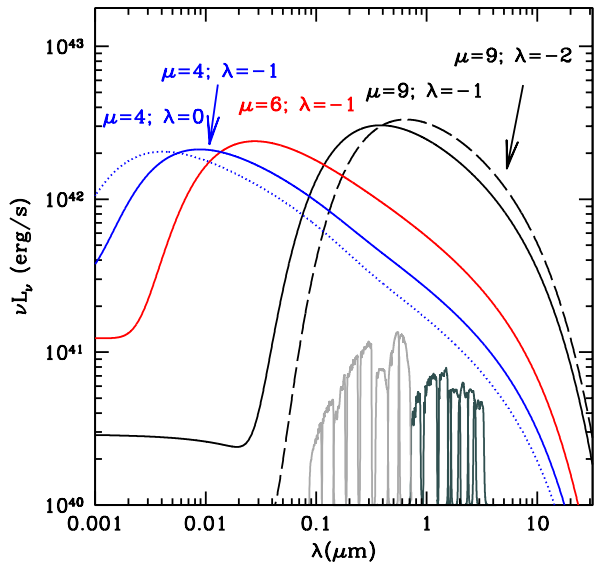}
\vspace{-1.4cm}
\caption{\footnotesize  Theoretical AGN spectra, normalized to the same total luminosity.  BH masses and Eddington ratios are marked on the figure. Continuum only. We report at the bottom of the figure the NIRCAM (light grey) and MIRI (dark grey) filter curves at $z=6$ restframe. 
\label{fig:spectra_agn}}
\end{figure}

\subsection{AGN spectra}
In this paper we  study the properties of BHs with masses and accretion rates that span a wide range, often covering regions of the parameter space that are far from the typical ``quasars" that are used as a benchmark to derive mean spectral energy distributions \citep[SEDs; e.g.,][]{1994ApJS...95....1E,2001AJ....122..549V,Richards06,2012ApJ...759....6E} that are then used for deriving fits for bolometric corrections \citep{Marconi2004,Hop_bol_2007,2012MNRAS.425..623L}.  See \cite{2016arXiv161108884J} for a first attempt at using observations to constrain AGN redshift spectral evolution. 

We therefore create theoretical AGN spectra that depend explicitly on the physical parameters, BH mass ($\mu$) and Eddington ratio ($\lambda$), inspired by the Shakura-Sunyaev solution \citep{Shakura1973}. The mass and accretion rate also determine the total luminosity,  $l=38.11+\lambda+\mu$.  In the classic Shakura-Sunyaev solution, the smaller the BH mass and the higher the Eddington ratio, the more the peak moves towards higher frequency: $T_{peak}\propto 10^{(\lambda-\mu)/4}$.  We follow here a variant based on the physical models developed by \cite{2012MNRAS.420.1848D}. Specifically, we calculate the energy of the peak of the SED as described in \cite{2016arXiv161105165T},  but adopt the default functional form of the spectrum used in Cloudy \citep{2013RMxAA..49..137F}:

\begin{equation}
f_\nu = \nu^{\alpha_{\rm UV}} e^{-\frac{h\nu}{kT_{BB}}}e^{-\frac{kT_{IR}}{h\nu}}+a \nu^{\alpha_x},
\label{AGN_cloudy_SED}
\end{equation}

with $\alpha_{\rm UV}$=0.5 and $\alpha_X=1$, $k T_{IR}=0.01$ Ryd.  Following \cite{2013RMxAA..49..137F}, the last term in Equation ~\ref{AGN_cloudy_SED} is set to zero below 1.36 eV (912 nm), and we do not extend the spectrum above 100 keV  (12.4~nm) restframe. The normalization of the X-ray component, $a$, is obtained through $\alpha_{OX}$, the exponent of the power-law connecting the continuum between 2 keV and 2500 \AA.  We assume that $\alpha_{OX}$ depends on mass and Eddington ratio as obtained using the models by \cite{2012MNRAS.420.1848D}, see, e.g., \cite{2012ApJ...761...73D}, and,  at difference with \cite{2013RMxAA..49..137F},  we include the contribution of the Big Bump, i.e., the pseudo black body in the first term of the right-hand side of Eq.~\ref{AGN_cloudy_SED}, to the emission at 2 keV, since for low-mass highly accreting BHs  the bump has a very high $T_{peak}$ and it may contribute to the X-rays. 

Examples for some masses and Eddington ratios are shown in Fig.~\ref{fig:spectra_agn}, where we have normalized the SEDs to the same luminosity to ease the comparison of the spectral shape. The SED is validated in Appendix  A against commonly used bolometric corrections and the SED used for their derivation. With this SED we can calculate monochromatic luminosities, as well as broadband ones, from infrared to hard X-rays. In principle, with an SED that depends on the BH physical properties, bolometric corrections become distributions, rather than a fixed bolometric correction at a given luminosity. We do not include attenuation for the AGN SED, although we include it statistically as an obscured fraction,  an approach that \cite{2017MNRAS.465.1915R} show produces a good match between UV and X-ray AGN LFs. We also do not include emission lines, which can boost both the AGN \citep{2001AJ....122..549V,2011ApJ...733...31H} and the galaxy  \citep{2010ApJ...708...26R,2011ApJ...743..121A,2013ApJ...763..129S,2014A&A...563A..81D} magnitude. Nebular emission will be studied in a companion paper.

We fit for the parameters describing the BH/AGN population, $\bar{\lambda}$, $\sigma_{\lambda}$ and $\sigma_{\mu}$ by minimizing the $\chi^2$ of the distance between the model and both the UV and  X-ray AGN LF.  For the UV AGN LF we consider the functional forms proposed by \cite{Willot2010} and \cite{2015ApJ...798...28K}, for the X-ray AGN LF, the upper and lower limits derived from a combination of LFs and upper limits from \cite{2016MNRAS.463..348V} and \cite{2016ApJ...827..150M}.

We include the luminosity range $\sim 8.8-11.8 \,{\rm L_\odot}$ in X-rays and $\sim 11.8-13.25 \,{\rm L_\odot}$ in UV, i.e., only the observed part of the LF, and we exclude the highest luminosities, as a duty cycle ${\cal D}=0.25$ is not appropriate for the most luminous quasars, which should have ${\cal D}\sim 0.75-1$. The set of parameters that best allows us to reproduce both the X-ray and UV LF is: $\bar{\lambda}=\log(0.40)$,  $\sigma_{\lambda}=0.40$,  $\sigma_{\mu}=0.50$.  Small variations on the best set are possible, but they do not change the results overall. The range of parameters and their uncertainties, as well as variations on the reference model are discussed in Appendix B.

\begin{figure}
\hspace{-.4cm}
\vspace{-1.4cm}
\includegraphics[width=3.5in]{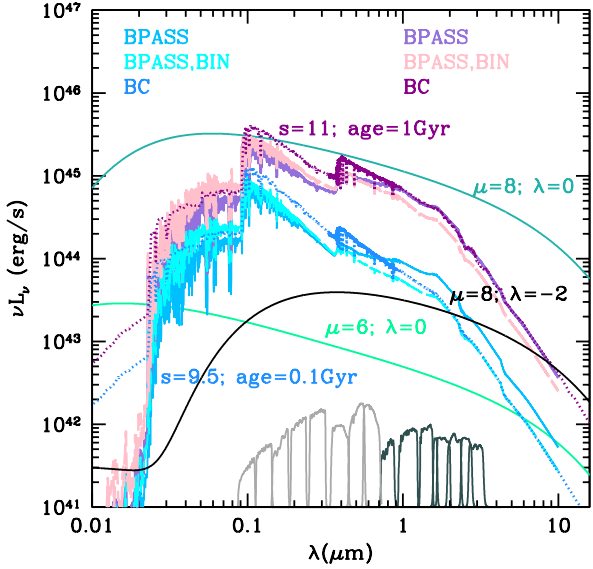}
\vspace{-1.4cm}
\caption{\footnotesize  Comparison between galaxy and AGN SEDs. Here the bolometric luminosity of the AGN is the same (black: $\mu=8$, $\lambda=-2$; turquoise: $\mu=6$, $\lambda=0$; blue-green: $\mu=8$, $\lambda=0$), the galaxy mass is $s=11$ (violet curves) and $s=9.5$ (blue curves) and the two galaxy spectra  are for  ${\rm Z}=0.2 {\rm Z_\odot}$ and different galaxy ages, using BC models  or BPASS models with and without binaries, normalized to the same stellar mass.  The chosen ages are the expectations from the assumption that the galaxies are on the main sequence, $\sim 0.1$ Gyr for $s=9.5$ and  $\sim 1$ Gyr for $s=11.5$. We report at the bottom of the figure the NIRCAM (light grey) and MIRI (dark grey) filter curves at $z=6$ restframe.
\label{fig:bias1}}
\end{figure}

\subsection{Galaxy spectra}
\label{galaxies}
Our starting point is the stellar mass and the redshift. To each galaxy we assign an SED from either Bruzual \& Charlot \citep[BC][version 2016]{2003MNRAS.344.1000B} or BPASS \citep{2009MNRAS.400.1019E,2016MNRAS.456..485S} models, using the minimum set of parameters that allows us to reproduce reasonably well the observed galaxy UV LF. We adopt a Salpeter initial mass function \citep{1955ApJ...121..161S} for consistency with most high-z studies. We assume constant star formation histories and map stellar mass to age through the galaxy main (or mass) sequence, connecting star formation rate (SFR) to galaxy stellar mass, as formulated for galaxies up to $z=6.5$ by  \cite{2015ApJ...799..183S}. Specifically, at $z=6$, $\log({\rm SFR})=0.54s-3.9$, with an intrinsic dispersion of 0.21 dex, and the age is obtained as the ratio of galaxy stellar mass to SFR.  As for most other relations and correlations adopted in this paper,  \cite{2015ApJ...799..183S} relation is determined for a subset of the mass/SFR range we are considering.  We then assign a galaxy to a metallicity bin, either $10^{-2.3}\, Z_\odot$, $10^{-0.7}\, Z_\odot$ or solar assuming a mass-metallicity relation.   The results are not strongly dependent on the metallicity grid, for the range of metallicity expected for galaxies at $z\sim 6$. Since observational constraints at $z=6$ are unavailable \citep[but see][for local analogs of high-redshift galaxies]{2017ApJ...834...51B}, we adopt the theoretical results from \cite{2016MNRAS.456.2140M}.  For completeness we have also performed a test based on the simple stellar populations of BC, but evolved with the code by 
\cite{2005MNRAS.362..799M}, and found results consistent with the standard BC models. Examples of galaxy and AGN SEDs are shown in Fig.~\ref{fig:bias1}.

\begin{figure}
\hspace{-.4cm}
\vspace{-1.4cm}
\includegraphics[width=3.5in]{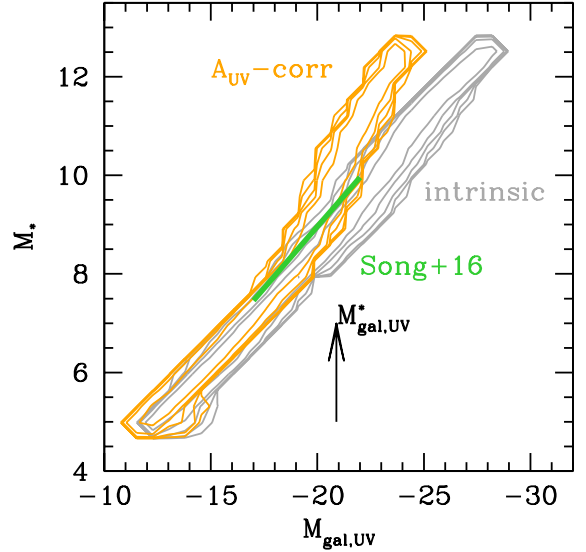}
\vspace{-1.4cm}
\caption{\footnotesize   Stellar mass vs UV magnitude. The grey contours show the intrinsic magnitude, the orange contours the attenuation-corrected magnitude, and the green line is the relation from \cite{2016ApJ...825....5S}. The arrow shows the location of the break in the galaxy luminosity function. 
\label{fig:UVcomp}}
\end{figure}

\begin{figure}
\hspace{-.4cm}
\vspace{-1.4cm}
\includegraphics[width=3.5in]{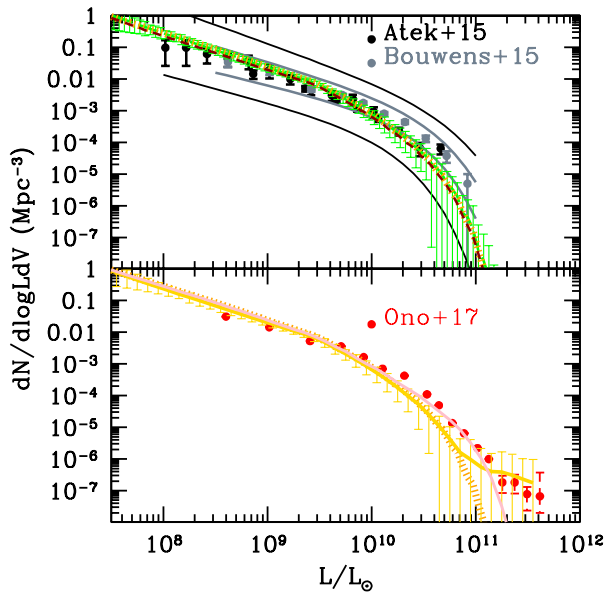}
\vspace{-1.4cm}
\caption{\footnotesize   Galaxy LF at 1550 \AA. Upper panel: In black and dark grey we show the error region of the functional forms of the LFs by \cite{2015ApJ...803...34B} and  \cite{2015ApJ...814...69A} The models are: dotted orange curve for BC, short-dashed dark red for single stars BPASS models, long-dashed green for BPASS models with binaries, where we show errorbars, that are similar for all other cases, but we do not include them for clarity. Lower panel: The yellow curve shows the LF we  obtain summing the galaxy and AGN light and should be compared to the red points, which represent the LF in \citet[][without correcting for AGN contamination]{2017arXiv170406004O}. The orange curve is reported from the upper panel to guide the eye. \label{fig:UVLF}}
\end{figure}

When we study the {\it intrinsic continuum emission} we adopt unattenuated spectra, and apply a magnitude-dependent attenuation \citep{1999ApJ...521...64M} only when estimating observable luminosities. We couple the relationship between extinction and UV slope, $A_{\rm UV}=4.43+1.99 \beta$ to the correlation between UV slope and magnitude in \citep{2014ApJ...793..115B} to obtain:
\begin{equation}
A_{\rm UV} = (0.58 \pm 0.57) - (0.67 \pm 0.28) (M_{\rm UV} + 19.5),
\label{eq:a_uv}
\end{equation}
where $M_{\rm UV}$ is the intrinsic, unattenuated magnitude obtained from the stellar population.  The errors are propagated ones and assumed to be uncorrelated, and no error on the slope of the relation between $A_{\rm UV}$ and $\beta$ is reported in the literature. This relation produces values similar to the model with no evolution in the relation between infrared excess and stellar mass of \cite{2016ApJ...833...72B}. We do not decrease $A_{\rm UV}$ below 0.5 in order to obtain a reasonable fit of the faint end of  the galaxy LF.  When we study the observable emission at wavelengths other than UV (1550 $\rm {\AA}$), we correct the galaxy luminosity assuming a dust extinction law \citep{2000ApJ...533..682C}, renormalized to obtain Eq.~\ref{eq:a_uv} at 1550 \AA. The uncertainty added in the model by extrapolating this and other relationships (e.g., stellar mass-SFR or mass-metallicity) to higher/lower galaxy masses and luminosities is difficult to assess, but we present most results in the following sections as a function of masses, magnitudes or luminosities, therefore the range of applicability can be inferred from the information given in this and the previous sections.    In Fig.~\ref{fig:UVcomp} we compare stellar masses and UV magnitude from the model to the relation proposed by \cite{2016ApJ...825....5S}. A better agreement at high-mass end would be obtained by limiting the galaxy age to $0.25$ Gyr.


With this set-up we have created galaxy LFs to compare with current observations \citep{2015ApJ...803...34B,2015ApJ...814...69A} and anchor our model (Fig.~\ref{fig:UVLF}). We obtain a reasonable match  around the knee of the galaxy LF for all  models, but we underestimate the bright end, at $L>2\times L_*$ and $L>4 \times L_*$ for the LF of  \cite{2015ApJ...814...69A} and \cite{2015ApJ...803...34B} respectively by $\sim 0.5$ dex, and the faint end, by $<0.35$ dex. The mismatch at the bright end is caused by an underestimate of the galaxy luminosity (once corrected for attenuation), while that at the faint end by overestimating the UV luminosity of low-mass galaxies. To improve the match with the faint end of the LF, we would need to either include only older stellar populations (age$\gta0.1$ Gyr) or set $A_{\rm UV}\geqslant1.5$, hinting at the presence of either non-star-forming galaxies, or galaxies with much higher levels of attenuation among low-mass galaxies. The former can be linked to low-mass galaxies being easily affected by stellar (and AGN) feedback, as well as from photoionization from the UV background.  Our model, based on the mass sequence of star-forming galaxies, does not include this putative population. The extinction we have included in our model is based on that of UV-selected galaxies, which by selection cannot be too dusty. Hints that dusty star forming galaxies exist at $z\sim 6$ are now seen in ALMA data \citep{2013Natur.496..329R,2016ApJ...832..114M,2017ApJ...842L..15S,2017Natur.545..457D}. {\it Hubble Space Telescope} observations, probing the restframe UV, are biased against quenched or dusty galaxies at high redshift, and {\it JWST}, with its optical/near infrared coverage will help unearth such galaxies. To improve the match at the bright end we should assume that galaxies with mass $>10^{10} \msun$ at $z=6$ are at most 0.25 Gyr old, which is consistent with the stellar ages of slightly less massive galaxies \citep{2013MNRAS.429..302C}, but at odds with the age of the stellar populations of some other observed galaxies \citep[e.g.,][]{2011MNRAS.414L..31R}.

\begin{deluxetable*}{ccccc}
\tablecaption{Reduced $\chi^2$ separately for the faint and bright end of the LF for models against the functional form of observational LFs. \label{key}}
\startdata
			& Bouwens+15 ($L<L_*$) & Bouwens+15 ($L>L_*$) & Atek+15 ($L<L_*$)  & Atek+15 ($L>L_*$) \\
			\hline 
BC   			& 4.11 				& 4.20				& 4.64$\times 10^{-2}$   & 0.17 \\
BPASS BIN 	& 3.09				& 3.96				& 4.21$\times 10^{-2}$	  & 0.11 \\
BPASS		& 2.21				& 5.19				& 2.45$\times 10^{-2}$   & 0.25 \\
\enddata
\end{deluxetable*}

\section{Results}

\subsection{Galaxy vs AGN: luminosities and biases}

Using the set-up described in the previous sections, we have in hand galaxy+AGN populations. We can therefore compare the properties of the two populations. We start with comparing the UV magnitudes\footnote{We calculate UV magnitudes as monochromatic magnitudes at 1550 \AA, while for NIRCAM and MIRI bands we convolve the SED with the filter response.} and we then move to the {\it JWST} bands, to study what type of AGNs can be more easily detected with future optical/near-IR observations.

\begin{figure}
\hspace{-.4cm}
\vspace{-1.4cm}
\includegraphics[width=3.5in]{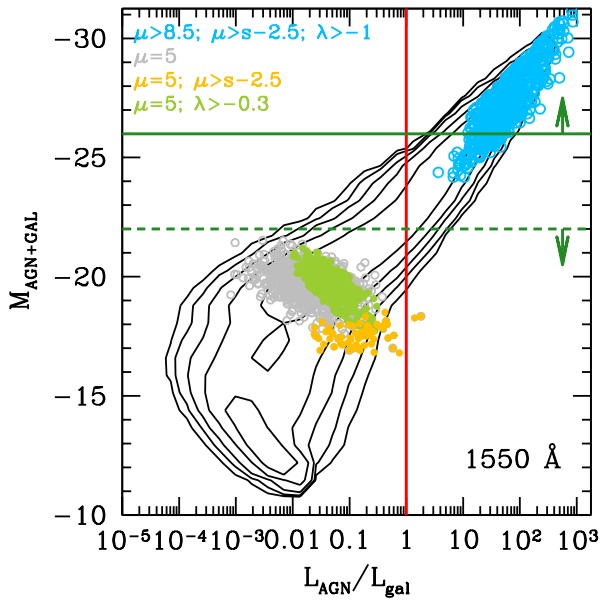}
\vspace{-1.4cm}
\caption{\footnotesize  Comparison between galaxy and AGN UV properties. The red line marks equal luminosity in the AGN and in the stellar population. The green solid line is the typical faint limit for quasar searches, while the dashed one marks the typical bright galaxies detected so far at $z\sim 6$. The cyan points show the typical physical  properties of the currently detectable quasars, i.e. above the flux limit and above the galaxy luminosity. The grey points mark galaxies hosting BHs with $4.75<\mu<5.25$ and the green and orange points are subsets of this population with either high accretion rates or large ratio between BH and galaxy mass, respectively. 
\label{fig:maggal_magAGN_UV}}
\end{figure}

\begin{figure}[t]
\hspace{-.4cm}
\vspace{-1.4cm}
\includegraphics[width=3.5in]{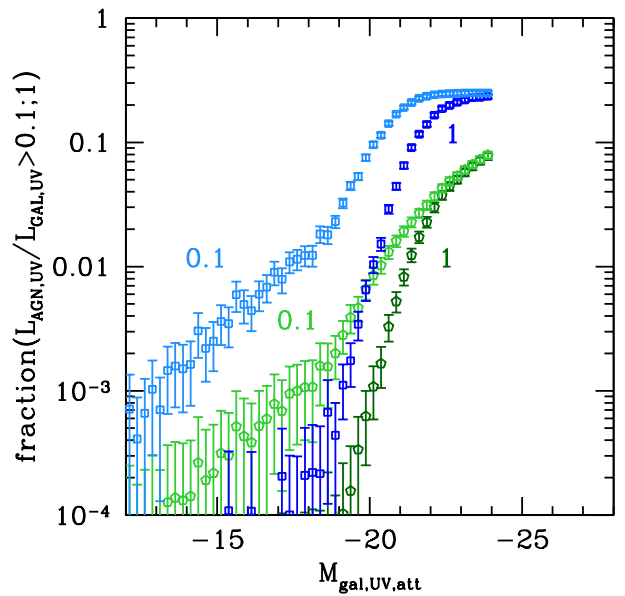}
\vspace{-1.4cm}
\caption{\footnotesize Fraction of galaxies hosting an AGN with luminosity $>0.1;~1$ times the galaxy luminosity.  Here the blue squares represent the intrinsic fraction, while the green pentagons correct for the AGN obscured fraction. Recall that by construction in our model only 25\% of galaxies host an active black hole, which is taken into account in this figure. 
\label{fig:agn_frac_UV}}
\end{figure}

\begin{figure}
\hspace{-.4cm}
\vspace{-2.25cm}
\includegraphics[width=3.5in]{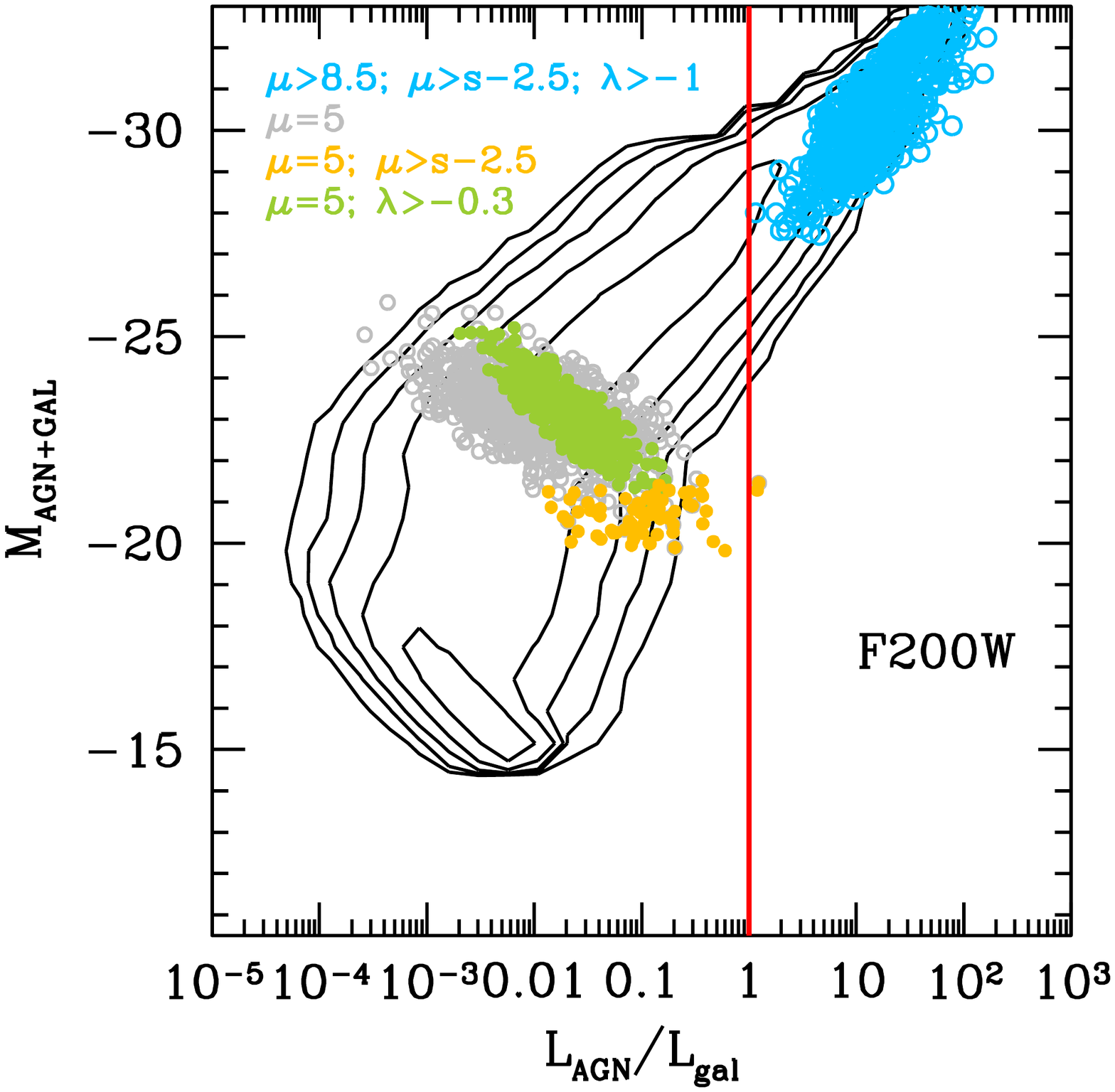}
\vspace{-1.4cm}
\hspace{-.4cm}
\includegraphics[width=3.5in]{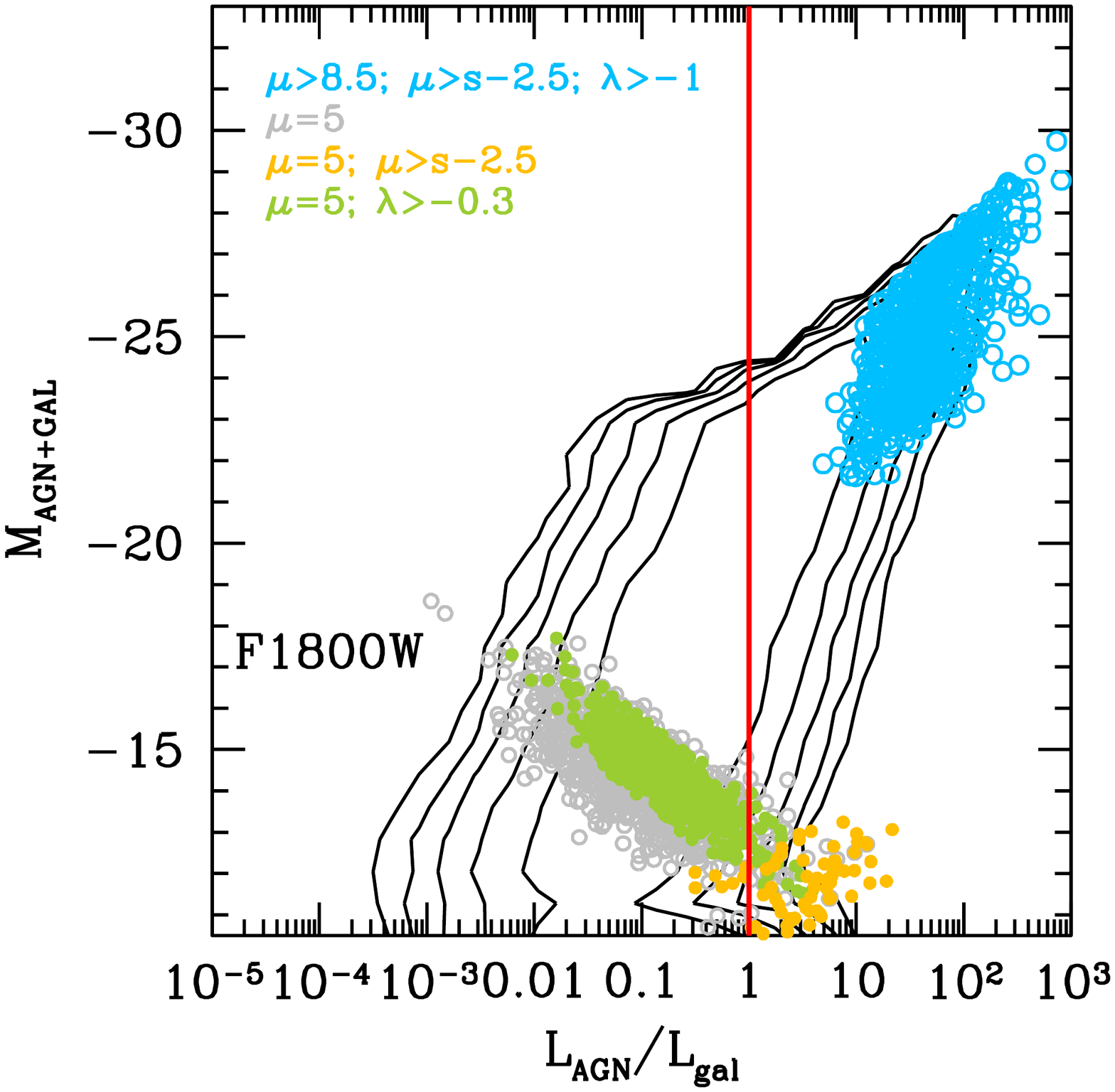}
\vspace{-1.4cm}
\caption{\footnotesize  Comparison between galaxy and AGN magnitudes in a NIRcam (F200W) and a MIRI band (F1800W). The red line is the 1:1 relation, and the points highlight the same populations as in Fig.~\ref{fig:maggal_magAGN_UV}. AGNs powered by small BHs are more easily identifiable at the reddest wavelengths.
\label{fig:maggal_magAGN_NIR}}
\end{figure}

In Fig.~\ref{fig:maggal_magAGN_UV} we compare  galaxy and AGN magnitudes in the UV (we adopt 1550 \AA~throughout, although most quasar studies use 1450 \AA~as a reference wavelength), where we have used Meurer-corrected BC models (results are qualitatively unchanged for all other stellar population models). We have not corrected the AGN luminosity for attenuation, making this an ``optimistic scenario". The currently known population of high-z quasars, detected in shallow surveys, is dominated by high-mass BHs with high accretion rates, and a large ratio between BH and galaxy mass. This bias is expected, as a fixed,  high-luminosity selection picks more frequently high-mass BHs hosted in low-mass galaxies than viceversa, because of the steep shape of the galaxy mass function  \citep{Shields2006,Lauer2007,2011MNRAS.417.2085V}. The population of the first BHs, possibly close to the seed mass are much harder to disentangle from the host galaxy, if it actively forms stars. We refer the reader to \cite{2016arXiv161005312N} for a dedicated study. 

Note also that by selecting sources with a total magnitude brighter than $-26$ (observations do not have an a priori AGN/galaxy separation), all sources are dominated by the AGN. This explains why no ``pure" galaxies have been identified in quasar searches that use as a starting point photometric information from large-shallow surveys. Conversely, since most galaxy searches are based on narrow-deep surveys, galaxies are typically fainter than  $-22$, and in this case the AGN contribution is small, in agreement with observations \citep{2013ApJ...778..130T,2015A&A...578A..83G,2016ApJ...823...95C} and theoretical models \citep{2016ApJ...820L...6V,2016arXiv160509394H}.  The recent surveys SHELLQ and GOLDRUSH, interestingly, select some of the most luminous galaxies as well as some of the faintest quasars at $z\sim 6$  \citep{2016ApJ...828...26M,2017arXiv170406004O,2017arXiv170405854M}. These surveys bridge the region where we predict galaxies and AGN/quasars coexist, and in fact \cite{2017arXiv170406004O} find a significant AGN contamination at  the bright end of the galaxy UV LF, see Fig. ~\ref{fig:UVLF}, in agreement with our model.

The fraction of galaxies hosting an AGN with luminosity larger than the galaxy UV luminosity, or 10\% of it, is shown in Fig.~\ref{fig:agn_frac_UV}, for the same model. Intrinsically, a fraction $\sim$ 20\% of galaxies with UV magnitude $\sim -21$ hosts an AGN with UV luminosity $>10\%$ of the galaxy, but taking into account the obscured fraction (type 1 vs type 2 AGNs) the detectable fraction decreases by about an order of magnitude. We show here the result obtained with BC models; they are statistically indistinguishable for BPASS. 

\begin{figure}
\hspace{-.4cm}
\vspace{-1.4cm}
\includegraphics[width=3.5in]{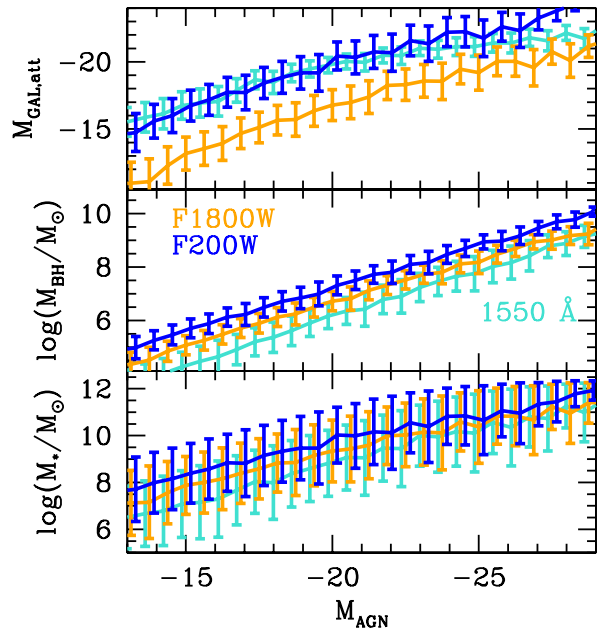}
\vspace{-1.cm}
\caption{\footnotesize  Properties of BHs and galaxies as a function of the AGN luminosity, weighted by number density. The x-axis is the non-attenuated AGN magnitude at various wavelengths, 1550 \AA (UV), a short-wavelength NIRCAM band (F200W) and a long-wavelength MIRI band (F1800W). The three panels show the logarithmic mean of galaxy stellar mass (bottom), BH mass (middle) and galaxy magnitude (top) and 1$\sigma$ dispersion for each quantity. 
\label{fig:properties}}
\end{figure}

Moving to the {\it JWST} bands, Fig.~\ref{fig:spectra_agn} shows that, since at high BH masses the spectrum peaks at redder frequencies, high mass BHs, with the SEDs peaking at $\sim 1 \mu m$ restframe, would  be favored over low mass ones, where the 1--5 $\mu m$ observer frame band at $z=6$ samples a region far from the peak of the SED. Additionally, given the inverse dependence of the peak energy and the Eddington ratio, at fixed BH mass the SED {\it shape} favors low Eddington ratios, but the SED {\it normalization}, via the total luminosity, favors high Eddington ratio. The latter is the most important of the two at fixed BH mass. However, at fixed total luminosity, proportional to $10^{\mu+\lambda}$, the shape of the SED is dominant in the {\it JWST} bands. For instance, the luminosity at 1--28 $\mu$m is higher for a BH with $\mu=8$ and $\lambda=-2$ than for a BH with $\mu=6$ and $\lambda=0$, although their bolometric luminosities are the same. In summary, at fixed {\it intrinsic bolometric luminosity} {\it JWST} will be biased towards detecting high-mass BHs, even with low intrinsic accretion rates, over low-mass ones. 

\begin{figure}
\hspace{-.4cm}
\vspace{-1.4cm}
\includegraphics[width=3.5in]{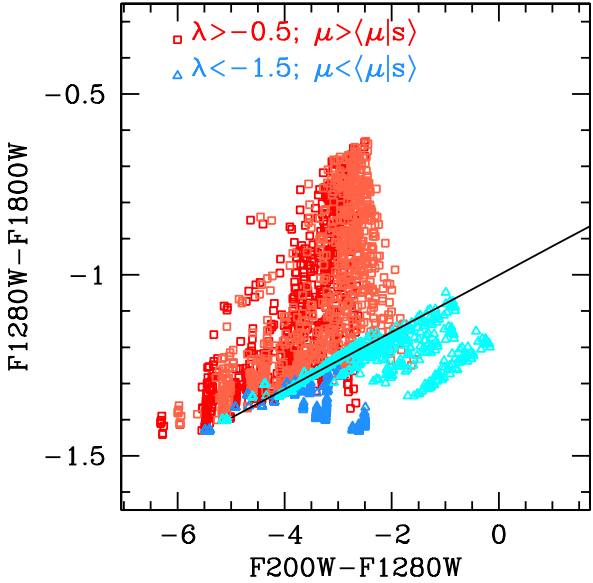}
\vspace{-1.4cm}
\caption{\footnotesize  Most promising color-color selection to separate AGNs from galaxies. The blue tracks show colors for pure AGNs, with different masses and accretion rates ($4\leqslant\mu\leqslant10$; $-2\leqslant\lambda\leqslant$, as marked in the figure). Recall that the peak temperature of the SED scales as $10^{(\lambda-\mu)/4}$ therefore different combinations of mass and accretion rates can produce the same color. The black curves show galaxy colors (younger to older from left to right) for all the stellar models used in this paper, without including attenuation, while the grey curves include a correction for attenuation as described in Section~\ref{galaxies}. The dark and light red squares show AGN-dominated galaxies, with and without attenuation respectively; the blue and cyan triangles show instead star-dominated galaxies with the same convention for attenuation. A color-color selection F1000W-F2100W vs F070W-F1000W gives similarly strong diagnostics.
\label{fig:colors}}
\end{figure}

However, we are interested in disentangling AGNs from their host galaxies. A more interesting assessment is on the relationship of AGN and galaxy as a function of physical properties, see for instance Fig.~\ref{fig:bias1}.  We are biased to identify low Eddington ratio BHs that are overmassive with respect to their host galaxy, e.g., $s=\mu+1.5$, with $\mu=8$, $\lambda=-2$ with respect to high Eddington ratio BHs which are ``normal" with respect to their host galaxy, e.g., $s=9.5$, with $\mu=6$, $\lambda=0$ in the example of the Figure.

\begin{figure}
\hspace{-.4cm}
\vspace{-1.4cm}
\includegraphics[width=3.5in]{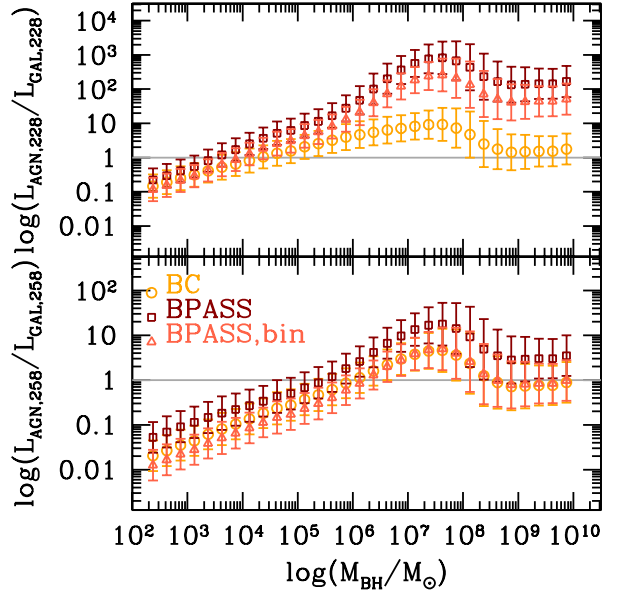}
\vspace{-1.4cm}
\caption{\footnotesize  Ratio of AGN and stellar contribution to the emission at 228 \AA~(54.4 eV) and 258 \AA~(47.9 eV). We compare BC models to BPASS models with and without binaries. 
\label{fig:line_ratio}}
\end{figure}

Finally, the contrast between AGN and galaxy increases at redder wavelengths, given the young ages of galaxies at $z\gtrsim 6$, therefore AGN searches will be favored at red wavelengths, as shown in Fig.~\ref{fig:maggal_magAGN_NIR} where we compare galaxy and AGN magnitudes in a NIRCAM band (F200W, the most sensitive for point sources) and in a red MIRI band (F1800W, more sensitive than F2100W or F2550W).  We note also that if one assumes AGN dust obscuration, the AGN-to-galaxy contrast will be even larger, enabling the use of shorter-wavelengths MIRI  filters. {\it JWST} can reach an absolute magnitude of about $-15$, with $-12$ reachable in the case of strong lensing clusters where we hope to gain 3 magnitude boost in some areas, provided we observe enough clusters.  We have used attenuated galaxy models, but we have not corrected the AGN luminosity for attenuation, making this an optimistic scenario from the point of view of AGNs.  

\begin{figure*}
\hspace{-.4cm}
\vspace{-2.25cm}
\includegraphics[width=\textwidth]{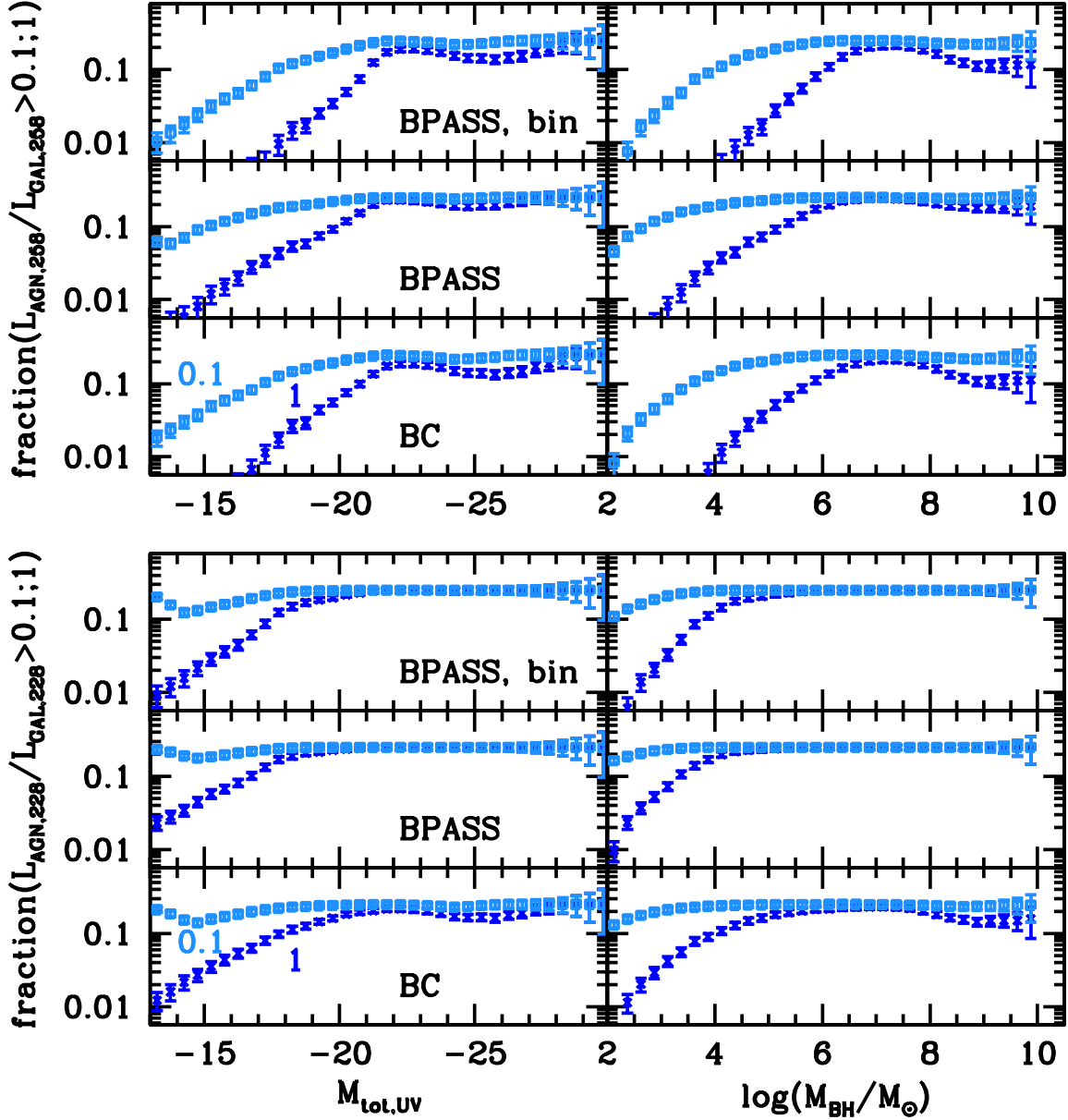}
\vspace{-1.4cm}
\caption{\footnotesize Fraction of galaxies as a function of total UV magnitude (galaxy+AGN, left) or BH mass (right) hosting an AGN with luminosity larger than a fraction $0.1;~1$ or  of the stellar luminosity at 228 \AA~(54.4 eV, \ion{He}{2}, bottom) or 258 \AA~(47.9 eV, \ion{C}{4}, bottom).  At a magnitude of $\sim -20$, the fraction of galaxies hosting an AGN contributing more than 10\% of the emission is $\sim 25$\% for \ion{He}{2} and 20\% for \ion{C}{4}. Recall that we assumed that only 25\% of galaxies host an active BH, therefore almost the totality of galaxies with an AGN have an AGN contribution $>10\%$ at these energies.
\label{fig:agn_frac}}
\end{figure*}

The relationship between BH, galaxy and their magnitudes is summarized in Fig.~\ref{fig:properties}. Here we calculate, in bins of non-attenuated AGN magnitude, the mean $\mu$, $s$ and galaxy magnitude, weighted on their number density. The latter takes into account the rarity of the most massive galaxies:  a quasar with $M_{UV}=-26$ has intrinsically a higher probability of being hosted in a galaxy with $s=12$, but such galaxies, if they exist, have a very low number density ($dN/d\log M\sim10^{-14} \,{\rm Mpc^{-3}}$ at $s=12$ using the mass function of Song et al. 2016), and therefore a small probability of being detected.   Weighting by the number density from the MF, which gives the probability of having a galaxy of a given mass in a given volume (i.e., a field at a given redshift), the weight shifts the expectation mass to lower mass, more common galaxies,  therefore making the most probable detectable host of a quasar with $M_{UV}=-26$ a galaxy with $s\sim10.5$. For $\rm {M_{AGN}}\sim -26$ at 1550 \AA, a reference value for the currently known population of $z\sim 6$ quasars, our models predicts that $s\sim 10.5$, $\mu\sim 8.5$ and $\rm {M_{\rm GAL,att}}\sim -21.5$. By including attenuation for the quasar magnitude using the same $A_{\rm UV}$ we use for galaxies, at an attenuated $\rm {M_{AGN}}\sim -26$ at 1550 \AA, then $s\sim 11$, $\mu\sim 9.3$ and $\rm {M_{\rm GAL,att}}\sim -22$. 

We have explored several color-color combinations that can be used to distinguish bona-fide AGNs from galaxies.   A possible distinction is  the presence of an actively accreting BH with a sizeable mass with respect to the stellar component, e.g., $\lambda>-0.5$ and $\mu>\langle \mu|s\rangle$. At the other end we envisage a slowly accreting BH with a low mass compared to the stars, e.g., $\lambda<-1.5$ and $\mu<\langle \mu|s\rangle$. In terms of AGN-to-stellar bolometric ratios, this distinction selects the upper cloud of the distribution, but it does not necessarily imply that $L_{\rm AGN}>L_{\rm GAL}$; the proposed cuts roughly correspond (in color-color space) to $L_{\rm AGN}>0.1L_{\rm GAL}$vs $L_{\rm AGN}<0.001L_{\rm GAL}$. With this practical definition we searched color-color combinations to find the most promising to disentangle galaxies with stronger or weaker AGNs. The best selection we found is a combination of the shortest wavelength NIRCAM filters (F090W to F200W;  F070W is also an option, but it has a much lower sensitivity) and the longest wavelength MIRI filters (F1800W, also in this case the choice is a compromise between clean selection and sensitivity, F2100W and F2550W having a much lower sensitivity but better discriminating power) with an intermediate-wavelength one, namely F1280W, F1000W or F770W \citep[see][for the latter choice]{2012ApJ...754..120M,2014A&A...562A.144M}. These combinations have the least amount of overlap between the two populations, and this is true in both cases where we correct for attenuation both galaxy and AGN or either one of them.  For instance, for the example shown in the Figure, using as a dividing line F1280W-F1800W$=0.08$(F200W$-$F1280W)$-0.99$, 15\% of the red (stronger AGN) points fall below the line, and 19\% of the blue (weaker AGN) points fall above the line. The fraction of misclassified sources is similar for the other combinations proposed above. In contrast, with combinations such as F200W-F444W vs F444W-F1800W or F200W-F1500W vs F1500W-F1800W, for instance, the fraction of misclassified sources is between 25\% and 50\%.  

As a proof of concept we corrected AGNs for reddening, simply using the same extinction curve used for galaxies, and found that the combinations above  remain the cleanest options. We report in Fig.~\ref{fig:colors} tracks of pure galaxies and AGNs, as well as a selection from our galaxy+AGN population matching the above criteria for F1280W-F1800W vs F200W-F1280W.   Results, however, could change once gas and dust are included (which will be explored more in paper II). 

\subsection{Galaxy vs AGN: UV radiation powering emission lines}
Some of the outstanding questions regarding the high-redshift Universe are whether AGNs or galaxies are responsible for reionization, and how many galaxies host an AGN.  We will address specifically the first in a companion paper, but in this study we start comparing galaxy and AGN emission in the UV, and specifically we focus on two energies, 54.4 eV and 47.9 eV (228 and 258 \AA). Photons with these energies or higher are required for producing \ion{He}{2} and \ion{C}{4} lines, now observed in high-z galaxy spectra \citep{2015MNRAS.450.1846S,2015MNRAS.454.1393S,2017MNRAS.464..469S,2017ApJ...836L..14M}. In Fig.~\ref{fig:line_ratio} we calculate the median of the ratio of the monochromatic luminosity produced by the AGN and produced by the galaxy stellar population, as a function of the BH mass.  This is the intrinsic luminosity, i.e., it does not include attenuation.

At 228 \AA~(\ion{He}{2}), the AGN contributes significantly, and sometimes dominantly, in galaxies with mass $8<s<11$ because these galaxies normally host BHs with mass $5<\mu<8$, where the spectrum has a strong UV component. At lower BH masses the spectrum becomes too hard, and at larger BH masses it becomes too soft (cf. Figs.~\ref{fig:spectra_agn} and~\ref{fig:bias1}). The AGN contribution is lower at 258 \AA~(\ion{C}{4}), but still significant, above 10\% for galaxies with mass $s\sim 10$ and BHs with mass $\mu\sim 7$. Caution should be taken in interpreting quantitatively these results, as stellar models in the far ultraviolet are not well calibrated and models are significantly different.  BC models do not tag the Wolf-Rayet phase in the tracks, but for those hot stars, they instead use the default planetary nebulae spectra from Rauch (2002), very close to blackbody spectra.  These may overestimate the flux of He ionizing spectra.  BPASS models instead include Wolf-Rayet stars;  these can be formed by mass transfer in binaries and not just the normal Conti mechanism. It is important, however, to draw attention to the possibility that AGN contribution in driving these lines could be substantial, even for young galaxies, and even including stellar binaries. 

The previous figures showed the emission contribution  when a galaxy contains an active BH, but, when estimating the fractional contribution of the AGN emission for a full population, we have to consider that only 25\% of galaxies are assumed to host an active AGN. Fig.~\ref{fig:agn_frac} takes this into account, showing as a function of attenuation-corrected galaxy UV magnitude (without including the AGN) the fraction of galaxies hosting an AGN with a luminosity larger than $0.1;~1$ times the galaxy luminosity at the same wavelength.  At a magnitude of $\sim -20$, the fraction of galaxies hosting an AGN contributing more than 10\% of the emission is $\sim 20-25$\% at 228 \AA~and $> 10$\% at 258 \AA.

\subsection{Luminosity functions and dependence on ``seed" BH properties}\label{sec:seed}
Our models are calibrated by requiring a good match with the region around the knee of the AGN LF in X-ray and UV. The faint end is an unexplored territory at $z\gtrsim 6$ and we investigate here the dependence on BH ``seed" properties, namely the minimum BH mass and the occupation fraction (OF) of BHs as a function of galaxy mass, as well as synergies with X-ray observations.

\begin{figure}
\hspace{-.4cm}
\vspace{-1.4cm}
\includegraphics[width=3.5in]{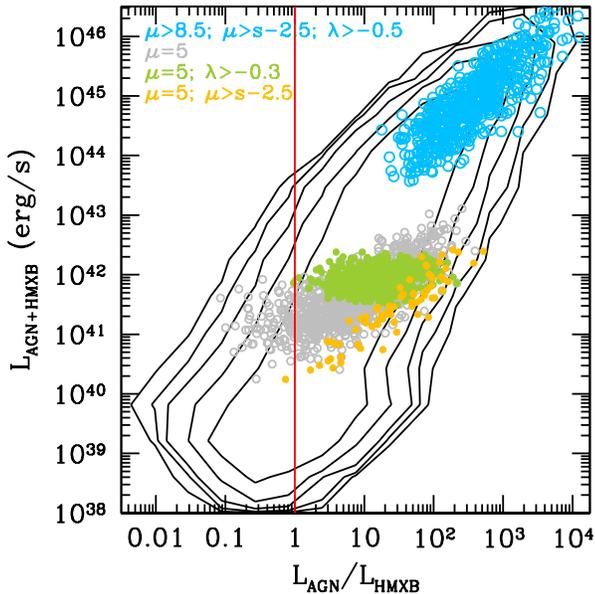}
\vspace{-1.4cm}
\caption{\footnotesize  Comparison between AGN luminosity vs luminosity in HMXBs in the host galaxy at 2-10 keV. Symbols as in Fig.~\ref{fig:maggal_magAGN_UV}.
\label{fig:LXRB_LAGN}}
\end{figure}

X-rays are usually considered a ``clean" way to select or confirm AGNs, as they are less contaminated by the host galaxy. The main source of confusion are high-mass X-ray binaries \citep[HMXBs, e.g.][]{2015ApJ...805...12L}, which are expected to be abundant in highly star-forming galaxies \citep{2012MNRAS.419.2095M}. The contamination, therefore, is likely to be more important in young, star-forming galaxies at $z\gtrsim 6$.  We assess the level of confusion by estimating the total luminosity of HMXBs as a function of SFR and redshift based on the scaling at 2-10 keV from model 269 in \cite{2013ApJ...764...41F}, shown by  \cite{2016ApJ...825....7L} to match very well observations up to $z\sim 2.5$, where good constraints from the X-ray data are available and  represent the best characterization of the scaling relations at all redshifts.   The ratio of HMXB emission to SFR in this model and in observations increases with redshift, therefore the difficulty of disentangling AGNs from HMXBs is increased at high redshift, $z\sim 6$, as evident in Fig.~\ref{fig:LXRB_LAGN}.

A comparison between Fig.~\ref{fig:LXRB_LAGN} and Fig.~\ref{fig:maggal_magAGN_NIR} shows that a significant fraction of faint AGNs, powered by low-mass BHs, can hardly be distinguished from the collection of HMXBs in the host galaxies, but this fraction is still much smaller than that where confusion from stellar population at optical/near-IR wavelengths is important, cf. the grey points in the two figures.

\begin{figure}
\hspace{-.4cm}
\vspace{-1.4cm}
\includegraphics[width=3.5in]{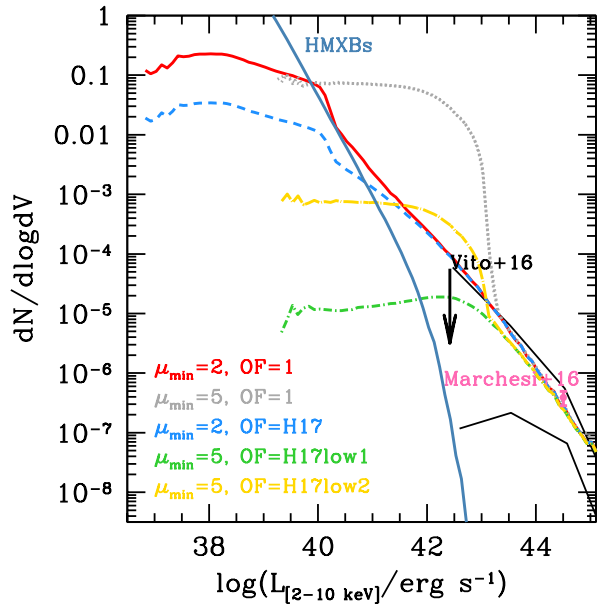}
\vspace{-1cm}
\caption{\footnotesize  AGN LF at 2-10 keV. The various curves explore different minimum BH mass or occupation fraction, as marked in the figure. The red and grey solid curves assume that every galaxy hosts a BH with minimum mass $\mu_{\rm min}=2$ and $\mu_{\rm min}=5$; the blue dashed curve assumes that the occupation fraction derived for ``light seeds" by \cite{2016arXiv160509394H}; the green short-dashed-dotted curve assumes $\mu_{\rm min}=5$ and the same functional shape of the occupation fraction, but shifted by a decade in stellar mass, and extended to zero (H17low1); the yellow long-dashed-dotted curve does not decrease the occupation fraction below 0.01 (H17low2). The black curves are upper and lower limits to the AGN LF from \cite{2016MNRAS.463..348V}, while the pink dot marks the point on the LF derived from the data in \cite{2016ApJ...827..150M}. The blue-green solid curve shows the LF of HMXBs, derived from model 269 in \cite{2013ApJ...764...41F} for the galaxy population in our model. On the left of this curve confusion by HMXBs would hamper identification of AGN. 
\label{fig:XLF}}
\end{figure}

\begin{figure}
\hspace{-.4cm}
\vspace{-1.4cm}
\includegraphics[width=3.5in]{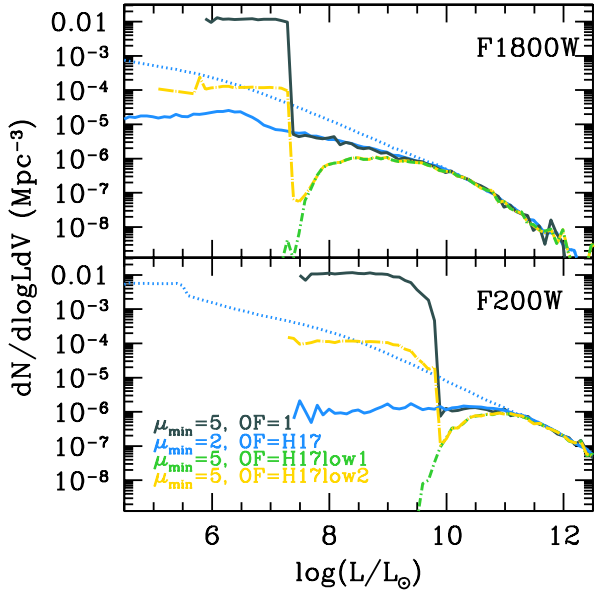}
\vspace{-1cm}
\caption{\footnotesize  AGN LF at F1800W and at F200W. Line styles are the same as in Fig.~\ref{fig:XLF}.  A model with a moderately high OF for massive seeds  (H17low2) shows a dip in the LF, when the galaxy outshines the AGN, but at the faint end the BH is as massive as the galaxy, and therefore the low occupation fraction is compensated by the AGN outshining the galaxy. When the occupation fraction is lower at low galaxy mass (H17low1) this does not occur and the faint end of the LF drops monotonically.
\label{fig:JWSTLF}}
\end{figure}

The same can be seen in the LF (Fig.~\ref{fig:XLF} and~\ref{fig:JWSTLF}), where selecting for AGNs with luminosity greater than the galaxy (the stellar population in optical/near-IR bands, HMXBs in X-rays), as a proxy for the population of sources more easily identifiable as AGN, significantly reduces the luminosity range that can be probed (compare dotted curves, all AGNs, with solid curves of the same color, uncontaminated AGNs). 

In Fig.~\ref{fig:XLF} we compare different models where we vary the minimum BH mass and the OF, based on theoretical models of BH formation \citep[see][for reviews]{2010A&ARv..18..279V,Haiman2012,2014GReGr..46.1702N}. Our basic model places a BH in every galaxy, with $\mu_{\rm min}=2$, a mass typical of ``light seeds", such as the remnants of Population~III stars and those formed by dynamical interactions in low-metallicity stellar clusters. In a variant, ``H17", we assume the occupation fraction derived by \cite{2016arXiv160509394H} in a dedicated cosmological hydrodynamical simulation (simulation ``D"):
\begin{eqnarray}
&{\rm OF}_{\rm D} = 1. - \frac{0.85}{1 + (\mu/\varepsilon)^{\beta}}\\
&\varepsilon=-0.077 (1+z) + 7.71\\
&\beta=2.30 (1+z)^{1.32}.
\end{eqnarray}
 Once the requirement that the AGN outshines the HMXBs is included, these two models are indistinguishable. 

To mimic the existence of more massive ``heavy seeds", we include models with $\mu_{\rm min}=5$, and either OF$=1$ (using the H17 occupation fraction results are identical to OF=1), or variations on ``H17" that take into account that the production of more massive BH seeds is rare and should occur in halos more massive than those where light seeds form \citep[][and references therein]{2016MNRAS.463..529H}. For this, we shift the functional form of the H17 OF by 1 dex in stellar mass, and modify the plateau at low mass with a linear extrapolation, down to OF$=0$ (``H17low1") or OF$=0.01$ (``H17low2").  We have compared the shifted scaling to the occupation fraction in two studies that model DCBHs, the simulation of \cite{2017MNRAS.470.1121T} and the semi-analytical model by \citep{2016MNRAS.462.2184H}. Occupation fractions are unpublished, but the authors provided us with their values, and the results are similar to the rescaled/shifted valued we estimated.
Significant differences between the models appear only at luminosities below $10^8 {\rm L_\odot}$, and there is degeneracy between seed mass and occupation fraction: more abundant light seeds are indistinguishable from rarer heavy seeds. 

\begin{figure}
\hspace{-.4cm}
\vspace{-1.4cm}
\includegraphics[width=3.5in]{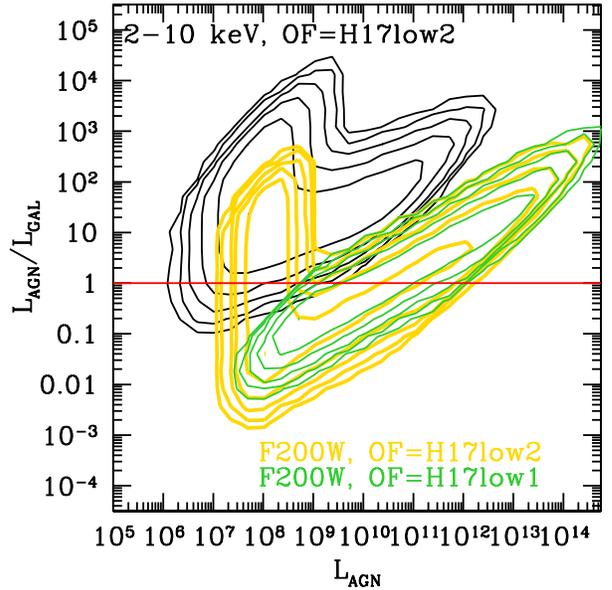}
\vspace{-1cm}
\caption{\footnotesize  Relation between AGN and galaxy luminosity at 2-10 keV for model H17low2 (black), as well as at F200W for models H17low2 (yellow) and H17low1 (green). The red line marks equal luminosity.  
\label{fig:lratio}}
\end{figure}

Having established the relevance of different assumptions, in Fig.~\ref{fig:JWSTLF} we limit the investigation to the three more physical models. We confirm the impression obtained from Fig.~\ref{fig:maggal_magAGN_NIR}: a red band enables a better discrimination from the host down to lower luminosities, but at the cost of a lower sensitivity: nominally, the point source detection limit is 8.6 $\mu$Jy for F1800W and 7.9 nJy for F200W for an exposure time of $10^4$ s and a signal-to-noise ratio of 10. The two effects approximately compensate. Fig.~\ref{fig:JWSTLF} shows also a non monotonic behavior of the LF for H17low2: this is caused by a combination of OF and relative luminosity between AGN and galaxy. In this model, at $s\sim9$ the fraction of galaxies with an AGN with luminosity larger than the galaxy becomes null, but, if the OF does not drop to zero, soon afterwards the average ratio of BH mass to galaxy mass increases, as the BH mass cannot go below $\mu=5$, and eventually the ratio between BH and stellar mass reaches unity at $s=5$. This population is a variant of the obese black holes proposed by \cite{2013MNRAS.432.3438A} and \cite{2016arXiv161005312N}. If instead the OF continues to decrease, such population does not contribute to the LF. The X-ray LF, instead, decreases monotonically because there is no galaxy mass at which the fraction of galaxies with an AGN with luminosity larger than the galaxy becomes null. The ratio between AGN and galaxy luminosity at 2-10 keV for H17low2, as well as at F200W for H17low2 and H17low1 is shown in Fig.~\ref{fig:lratio} to exemplify the arguments above.

\section{Conclusions}
We have presented a population synthesis model for galaxies, black holes and AGNs at high redshift, primarily towards the faint end of the LF. In this first paper we have focused on the description of the method, its validation against observational constraints, and the analysis of the relative properties of  galaxies and AGNs, in the restframe UV and X-rays, where observations are available, and in the restframe optical/near-IR, where observations will be soon available owing to {\it JWST}. 

The model assumptions are as follows:
\begin{itemize} 
\item The model builds a galaxy/AGN population taking as starting point the galaxy MF. 
\item Galaxies are characterized by an SED determined by age and metallicity. The former is assigned by assuming a relation between stellar mass and SFR (main sequence) and the latter from a mass-metallicity relation.  
\item BHs have a mass that scales with the galaxy stellar mass, and a luminosity given by a log-normal distribution of Eddington ratios, with parameters fitted by requiring a good match with upper/lower values to the AGN LF at $z=6$. The AGN SED depends on the physical properties of BHs, namely their mass and Eddington ratio. 
\end{itemize}

The main results are as follows:
\begin{itemize} 
\item For high-redshift galaxies, with stellar ages $<1$ Gyr, confusion between the galaxy and the AGN is higher at UV and blue optical wavelengths, where uncertainties in dust attenuation are also significant. 
\item We propose a color-color selection, e.g., F1280W-F1800W vs F200W-F1280W,  to separate galaxies with stronger and weaker AGN in {\it JWST} photometric observations. 
\item We estimate the AGN contribution at the energies driving \ion{C}{4} and \ion{He}{2}. At a magnitude of $\sim -20$, the fraction of galaxies with an AGN contributing more than 10\% to driving the \ion{He}{2} line is $\sim 20-25$\%, and $\sim 10-20\%$ fraction for \ion{C}{4}.
\item We adopt recent determinations of the redshift evolution of the relation between SFR and HMXB luminosity to predict where ``stellar contamination" affects X-ray observations, and establish a baseline for multi-wavelength studies. 
\item For realistic assumptions, the faint end of the X-ray and UV to near-IR LF does not depend appreciably on the minimum BH mass and on the fraction of galaxies hosting a BH, especially considering the degeneracies between these parameters. The difficulty of distinguishing the AGN emission from starlight and HMXBs at low AGN luminosity hinders a clean distinction between these properties.
\vspace{1cm}
\end{itemize}

\acknowledgements 
We are  grateful to the reviewer for her/his suggestions and careful reading of the manuscript. MV thanks heartily Matt Lehnert for helping her unravel the mystery of magnitudes, ``a quaint unit of historical interest", cit.~Cloudy \& Associates (www.nublado.org), Alice Shapley for thoughtful conversations and her kind help with galaxy spectra, Roberto Gilli and Roberto Decarli for constructive discussions and comments on the manuscript and F. Vito and B. Lehmer for help with HMXBs. MV and MT acknowledge funding from the European Research Council under the European Community's Seventh Framework Programme (FP7/2007-2013 Grant Agreement no.\ 614199, project ``BLACK'').  AER is grateful for the support of NASA through Hubble Fellowship grant HST-HF2-51347.001-A awarded by the Space Telescope Science Institute, which is operated by the Association of Universities for Research in Astronomy, Inc., for NASA, under contract NAS 5-26555.

\bibliography{../../../biblio_complete}

\appendix
\section{{Validation of the AGN SED}}
In this Appendix we validate the AGN SED that we created against commonly adopted SEDs derived from observations. An important consideration is that our AGNs span a large range in mass and accretion rates, while the observed population of quasars samples a biased region of the $\mu$-$\lambda$ parameter space. 

\cite{2011ApJ...728...98D} find no evidence of a dependence of the ratio of optical to bolometric luminosity with BH mass \citep[see also][]{1999PASP..111....1K}, and it seems clear that a ``basic" accretion disk model fails to reproduce all the features of observed spectra. However, we argue in the following that our templates provide a qualitative good approximation to SEDs in the range of masses and accretion rates probed by observations, at a level sufficient for the scope of this study. The assessment below shows that our approach produces reasonable results within a physically-motivated framework.

\begin{figure}[!h]
\hspace{-.4cm}
\vspace{-1.4cm}
\includegraphics[width=3.5in]{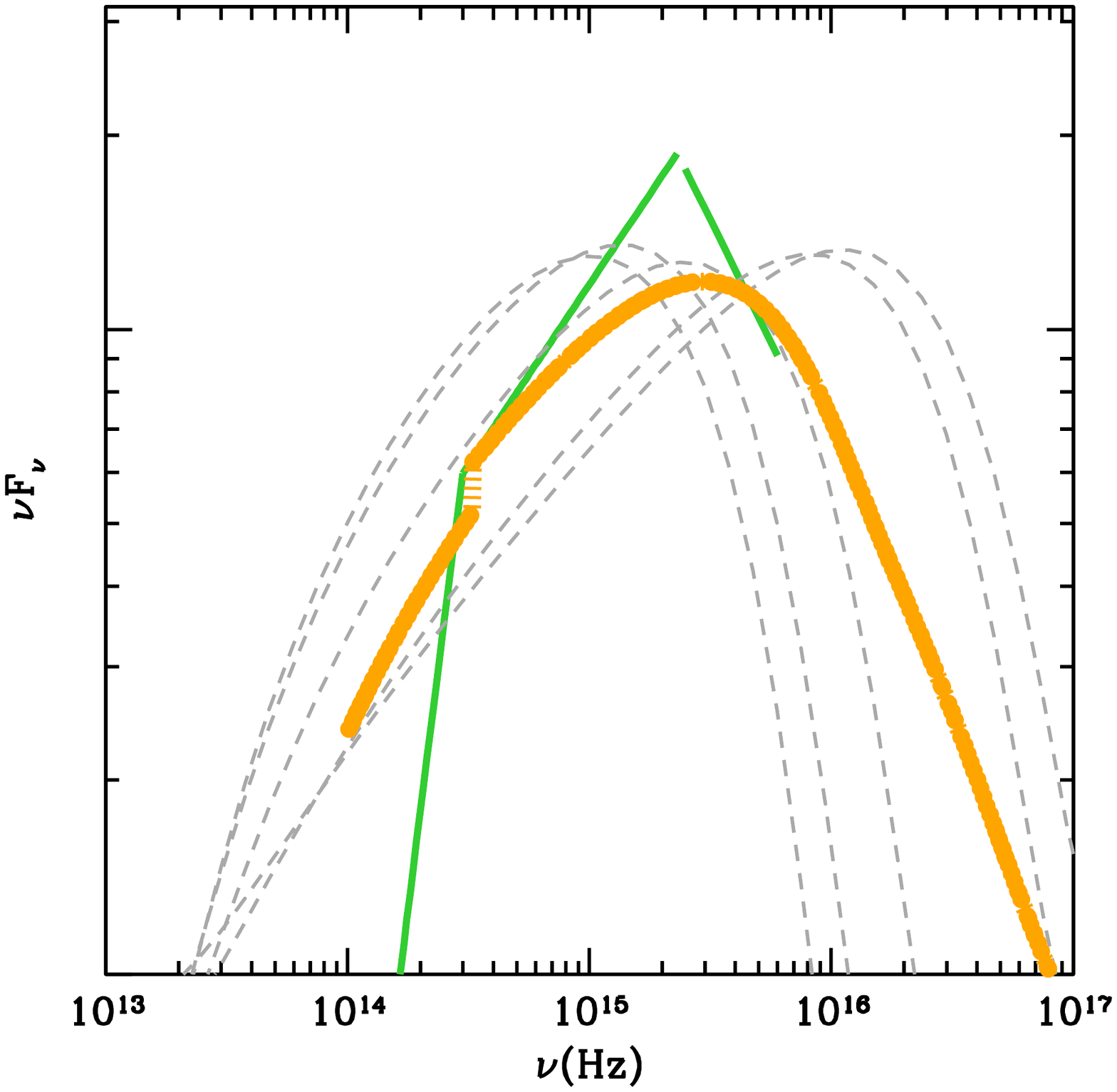}
\vspace{-1.cm}
\caption{\footnotesize Comparison between the adopted AGN SED with the template spectrum used by \cite[][green broken power-laws]{Marconi2004}. The short-dashed grey curves show, from left to right, the SEDs of BHs with:  $\mu=9$, $\lambda=-0.52$; $\mu=8$, $\lambda=-0.52$; $\mu=8.5$, $\lambda=-0.52$; $\mu=7.5$, $\lambda=0$; $\mu=7$, $\lambda=0$. The orange dotted curve is  the geometric mean of a suite of SEDs generated for a population of BHs with $\bar{\lambda}=\log(0.40)$,  $\sigma_{\lambda}=0.40$, $\sigma_{\mu}=0.50$,  and ${\cal D}=0.25$, selecting only AGNs with $\lambda>-1$ and bolometric luminosity $>10^{44}\, {\rm erg\,s^{-1}}$. To obtain a sharper cut-off at red wavelengths, we could increase $\tau_{IR}$, set to $k_B T_{\rm IR} = 0.01$ Ryd in the default spectrum.
\label{fig:AGN_SED_marconi}}
\end{figure}

\begin{figure}
\hspace{-.4cm}
\vspace{-1.4cm}
\includegraphics[width=3.5in]{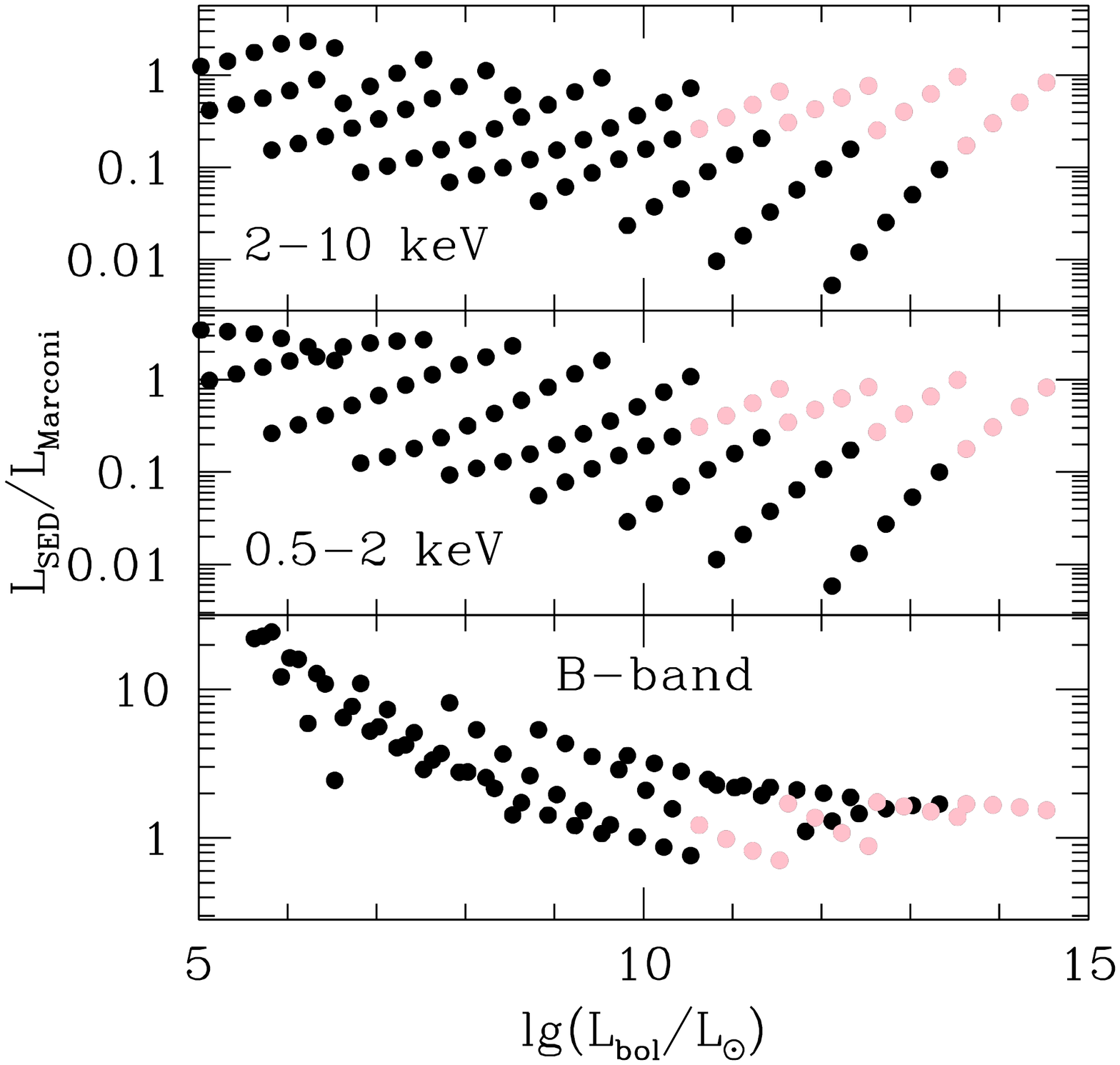}
\includegraphics[width=3.5in]{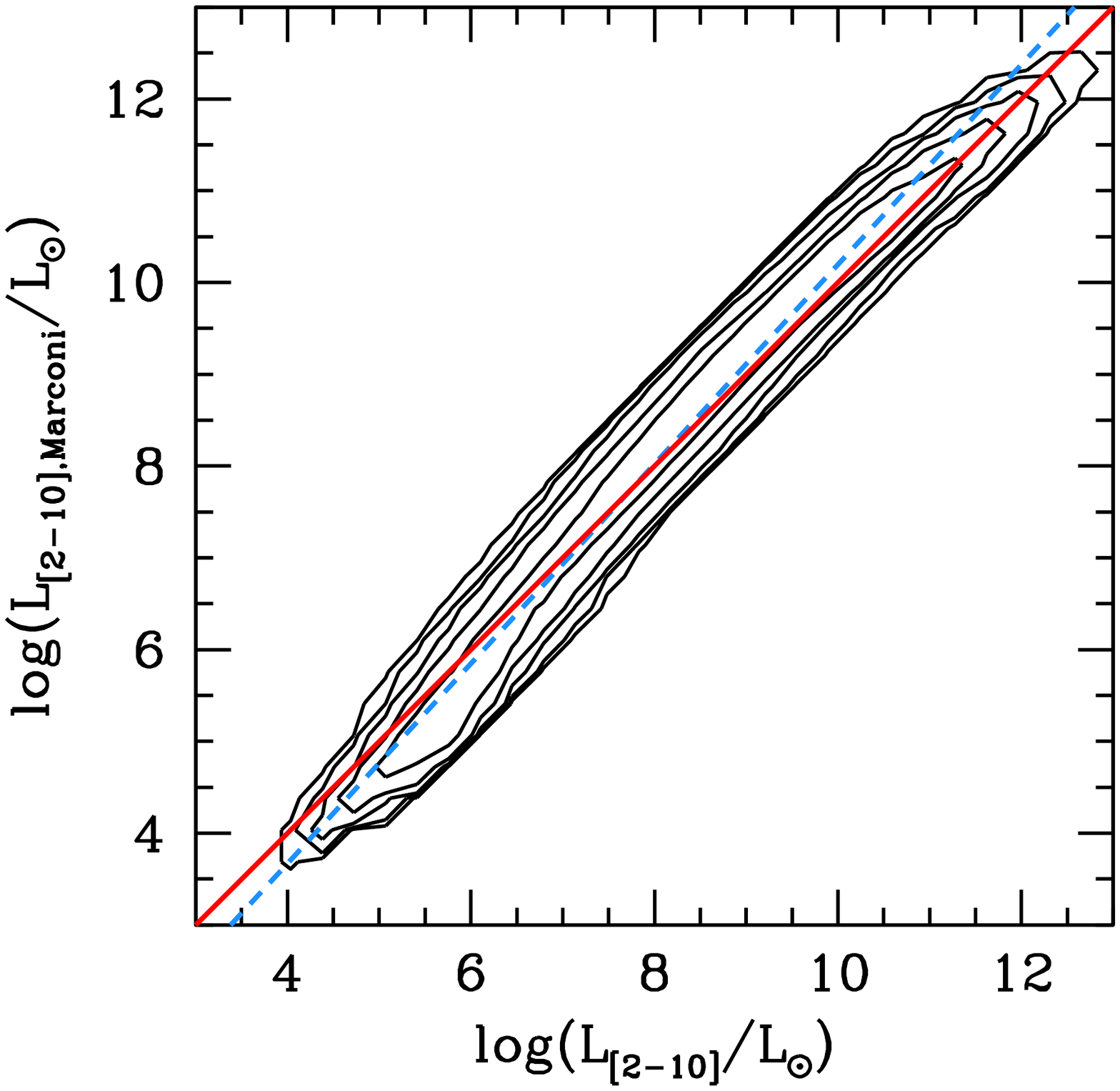}
\vspace{-1.cm}
\caption{\footnotesize  Left: Bolometric corrections obtained for our AGN SED compared to those by \cite{Marconi2004}.  The pink points highlight BHs with $\mu\geqslant7$ and $\lambda\geqslant-1$, corresponding to the typical masses and accretion rates for quasars, for which the standard SEDs have been calculated. Right: Comparison of the distribution of X-ray luminosities at 2-10 keV calculated from our SEDs to those obtained from  \cite{Marconi2004} for a population of BHs with $\bar{\lambda}=\log(0.40)$, $\sigma_{\lambda}=0.40$ and $\sigma_{\mu}=0.50$. The red solid line is the 1:1 relation, while the blue dashed line shows a least square fit in log space (slope: 1.1, intercept: -0.7).
\label{fig:bolcorr1}}
\end{figure}

In Fig.~\ref{fig:AGN_SED_marconi} we compare the shape of our SED to the combination of broken power-laws adopted by \cite{Marconi2004} and later by \cite{Hop_bol_2007}. The shape and location of the peak match well when we select only BHs with properties that correspond the general sample of observed quasars, i.e., with high BH mass and accretion rate. It is particularly encouraging that the geometric mean of the SEDs  for a population of BHs with the reference parameters we used, $\bar{\lambda}=\log(0.11)$,  $\sigma_{\lambda}=0.30$, $\sigma_{\mu}=0.75$,  and ${\cal D}=0.25$, is in good agreement with the spectrum template.

More in detail we can appreciate a comparison with the bolometric corrections in the standard reference bands, B-band, 0.5-2 keV, and 2-10 keV, using the same conventions as in  \cite{Marconi2004}, i.e.,  $L_{\rm bol}/\nu_B L_{\nu_B}$,  $L_{\rm bol}/L_{\rm 0.5-2keV}$ and  $L_{\rm bol}/L_{\rm 2-10keV}$. The comparison is shown in Fig.~\ref{fig:bolcorr1}, left.   Each diagonal sequence is for a different BH mass from $\mu=2$ to $\mu=10$, left to right. In the B-band, for each sequence the Eddington ratio increases from top to bottom in each sequence, from $\lambda=-2.7$ to $\lambda=0$. In X-rays, instead, the Eddington ratio decreases from top to bottom, with the same $\lambda$ range.  Compared also to the recent estimates of bolometric corrections by \cite{2012MNRAS.425..623L}, the agreement in the B-band, for the same luminosity range probed by observations, $10^{10}<L_{\rm bol}<10^{12} \, {\rm L_\odot}$, is very good, while our X-ray bolometric corrections are higher.  In both the comparison with \cite{Marconi2004} and  \cite{2012MNRAS.425..623L}, the underestimation of the X-ray luminosity appears to be caused by not having included a reflection component on top of the power-law in our SED. When included in a statistical sample, with the population properties with the reference parameters we used, the differences are minimized, as shown in the right panel of Fig.~\ref{fig:bolcorr1}, since most BHs are accreting at relatively high rates ($\bar{\lambda}=\log(0.11)$). We have checked that this difference does not affect significantly the X-ray LF.

\begin{figure}[t]
\hspace{-.4cm}
\vspace{-1.4cm}
\includegraphics[width=3.5in]{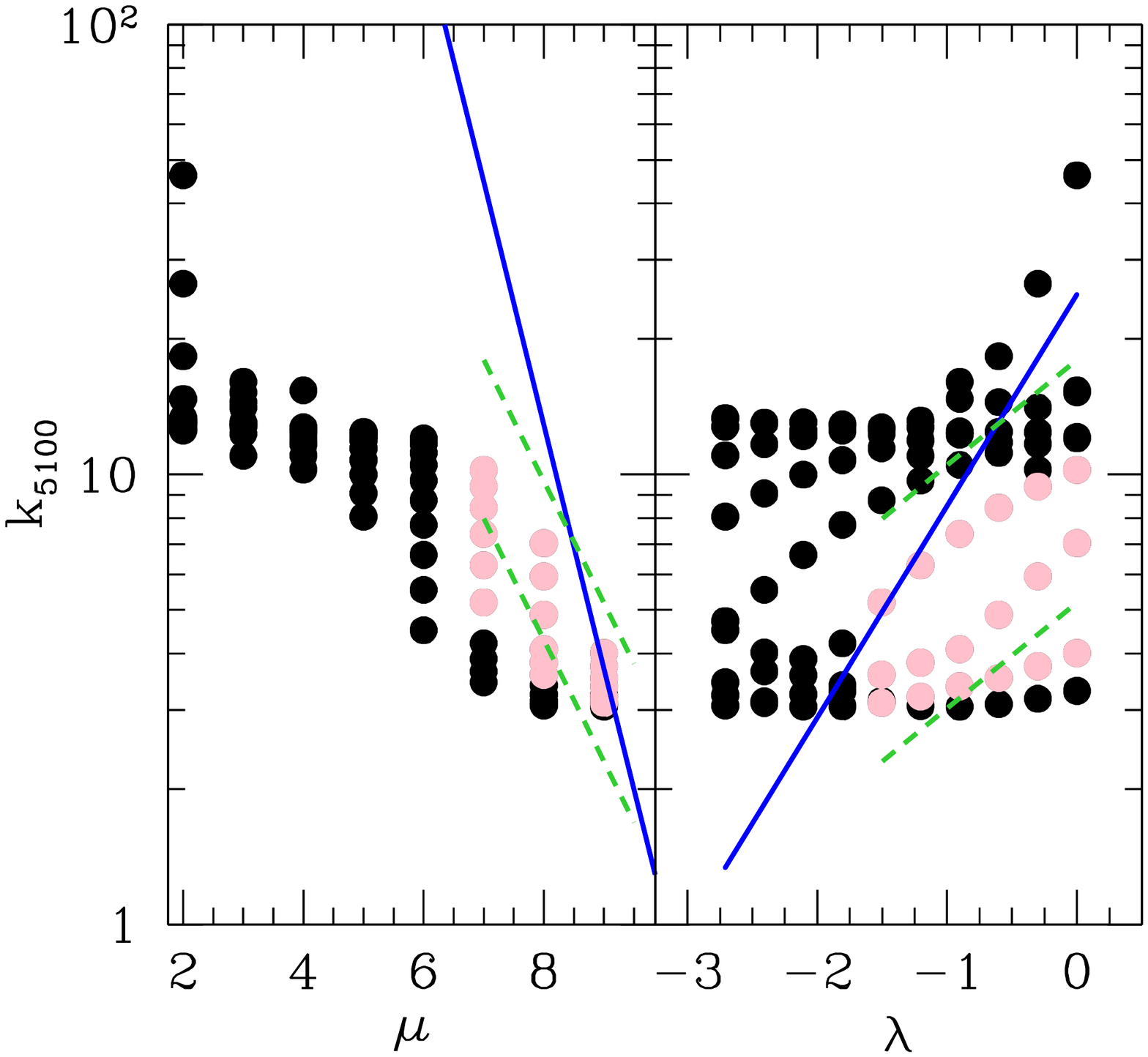}
\vspace{-1.cm}
\caption{\footnotesize Bolometric correction, $k_{5100}$, defined as $L_{\rm bol}/L_{\rm 5100}$, where $L_{\rm 5100}$ is the monochromatic luminosity at 5100 \AA~ vs BH mass, $\mu$, or Eddington ratio, $\lambda$. The black points are obtained from our SED,  with the pink points selecting BHs with $\mu\geqslant7$ and $\lambda\geqslant-1$. The blue line reports the correlations obtained by \cite{2012MNRAS.425..907J} for $k_{5100}$ vs $\mu$ or $\lambda$ separately. The green dashed lines show the joint $\mu$ and $\lambda$ dependence obtained by combining the two separate fits, these lines are shown only for BHs with $-1.5<\lambda<0$, $7\leqslant \mu \leqslant 9$ to match the mass-accretion range probed by observations.
\label{fig:bolcorr2}}
\end{figure}

In Fig.~\ref{fig:bolcorr2} we compare bolometric corrections at 5100 \AA~ with the results by \cite{2012MNRAS.425..907J}, where they fit the dependence from mass and accretion rate on a large sample of nearby unobscured AGNs. In the left panel, the accretion rate increases bottom to top, while in the right panel there is a more complicated behavior but, overall mass increases from top to bottom. Interestingly, while the separate fits for mass and Eddington ratio from \cite{2012MNRAS.425..907J} are not in good agreement with our model, the joint $\mu$ and $\lambda$ dependence obtained by combining the two separate fits matches well our SED values. 

\section{Best fit model parameters, uncertainties and variations}
\begin{figure}[t]
\hspace{-.4cm}
\vspace{-1.4cm}
\includegraphics[width=3.5in]{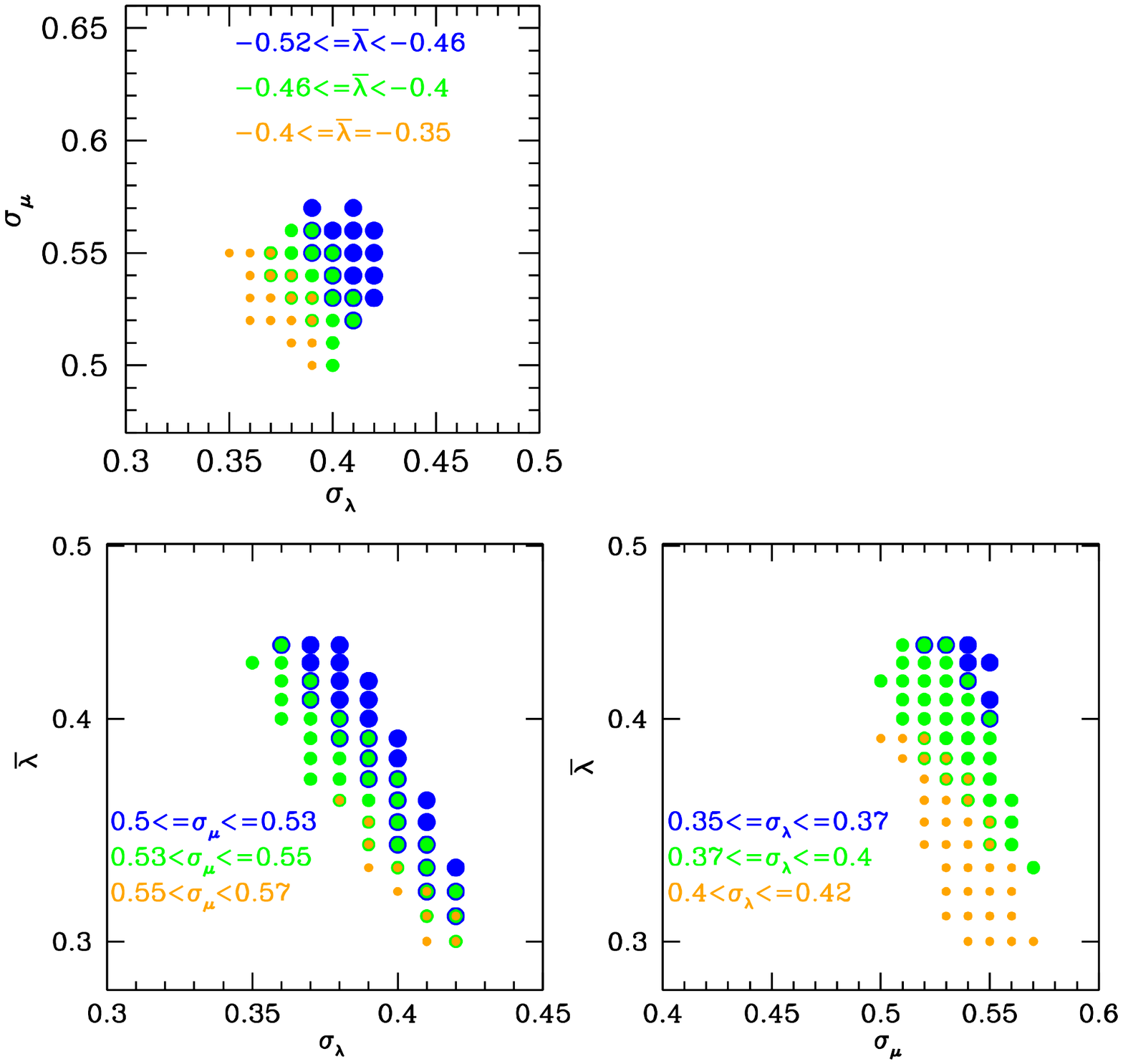}
\vspace{-1.cm}
\caption{\footnotesize Range of parameters providing a goodness of fit comparable to the set of parameters that best allows us to reproduce both the X-ray and UV LF. In each panel we show how 2 of the 3 parameters vary as a function of the third. Parameters are correlated: a smaller $\bar{\lambda}$ requires larger $\sigma_{\lambda}$ and $\sigma_{\mu}$. 
\label{fig:errors}}
\end{figure}

In Fig.~\ref{fig:errors} we show the parameters giving $\chi^2<0.35$ and how they vary in a correlated way. All these models provide similar results for the conclusions of the paper. The three free parameters are not independent. A smaller $\bar{\lambda}$ can be accommodated with larger $\sigma_{\lambda}$ and $\sigma_{\mu}$, and vice-versa. The range reported is derived from a grid of values, and for combinations of the parameters within the range provided, the chi-squared value is within a similar range.

We discuss here a super-maximal case, where we fit the parameters to the UV LF by Giallongo et al. (2015). This exercise requires a different approach. 
The AGNs are detected in X-rays, but the LF is provided in the UV, for the total UV luminosity, without separating galaxy from AGN. In order to mimic the same approach, we calculate the LF by applying only a correction for Compton Thick AGNs (to reproduce the X-ray selection) and by including both AGN and galaxy luminosity when calculating the LF in UV (to reproduce the total UV luminosity used by Giallongo et al. 2015 to estimate the AGN LF).  We are unable to find an acceptable fit for this LF for our standard BH-galaxy relationship, although, if we limit the comparison only to the magnitude range of the data of Giallongo et al. 2015 (from $-19$ to $-21$), i.e., we do not try to fit the full LF,  the  model parameters of our reference case, produce a LF compatible with that of Giallongo et al. 2015. We confirm the results by \cite{2017arXiv170304895Q} that the UV luminosity is dominated by the galaxy stellar population in this magnitude range, with the AGN-only luminosity function, that is, without including stars in the source luminosity,  $\times 0.04$  what we obtain including both stellar and AGN light in the ``AGN" luminosity. 

\begin{figure}
\hspace{-.4cm}
\vspace{-1.4cm}
\includegraphics[width=3.5in]{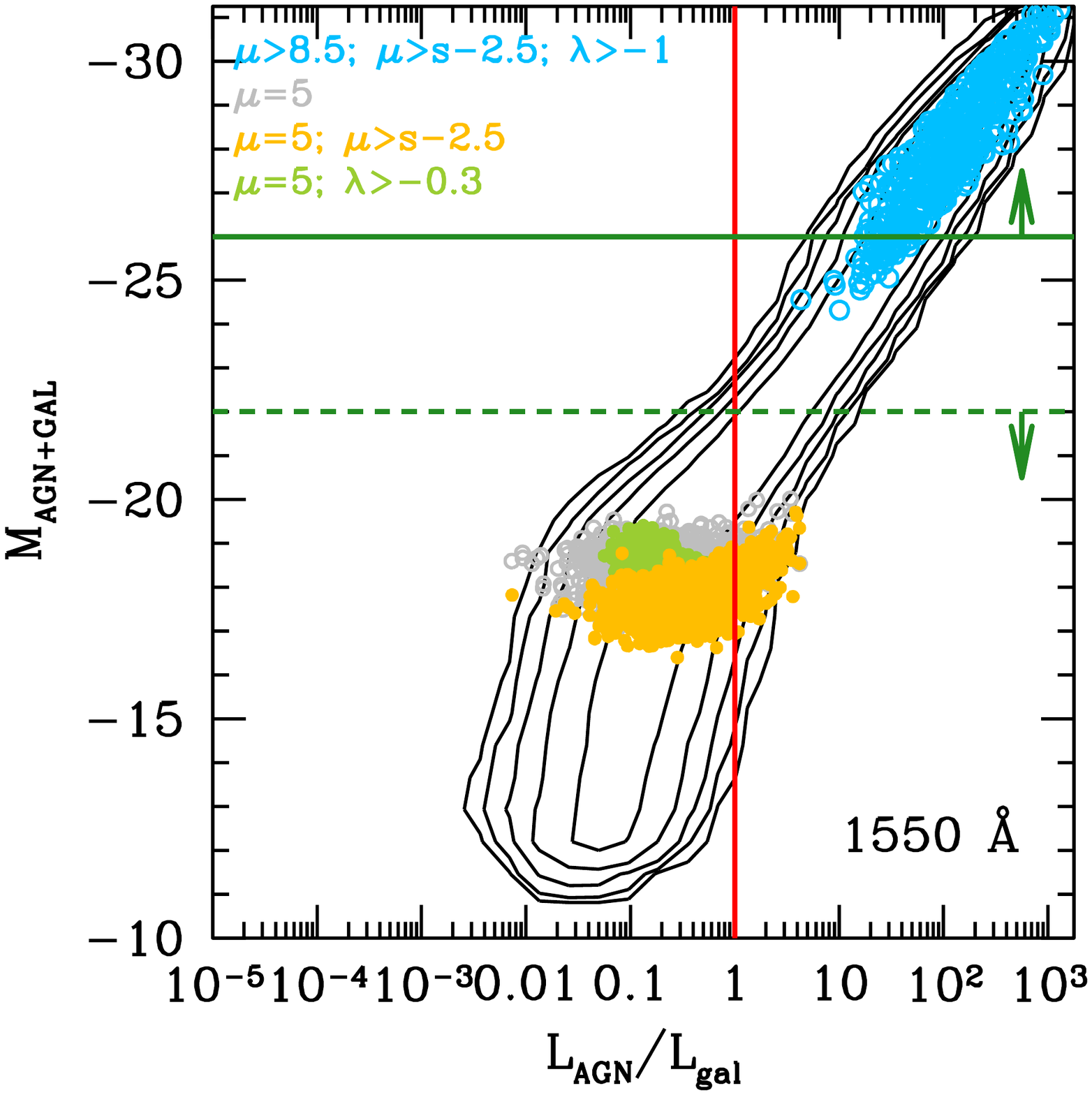}
\includegraphics[width=3.5in]{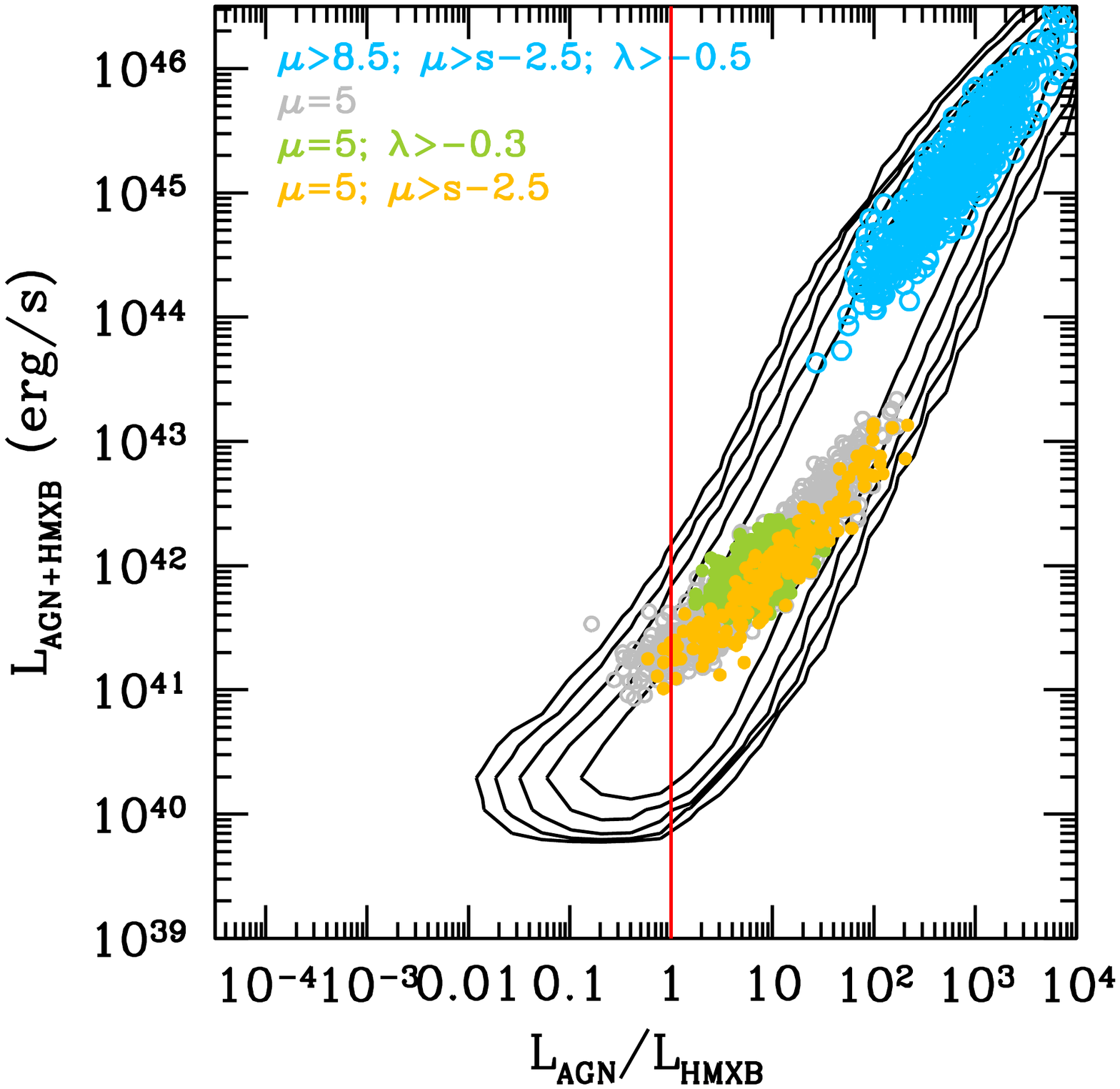}
\vspace{-1.cm}
\caption{\footnotesize Compare to Figures~\ref{fig:maggal_magAGN_UV}, \ref{fig:LXRB_LAGN}. 
\label{fig:contamination_max_G}}
\end{figure}

\begin{figure}
\hspace{-.4cm}
\vspace{-1.4cm}
\includegraphics[width=3.5in]{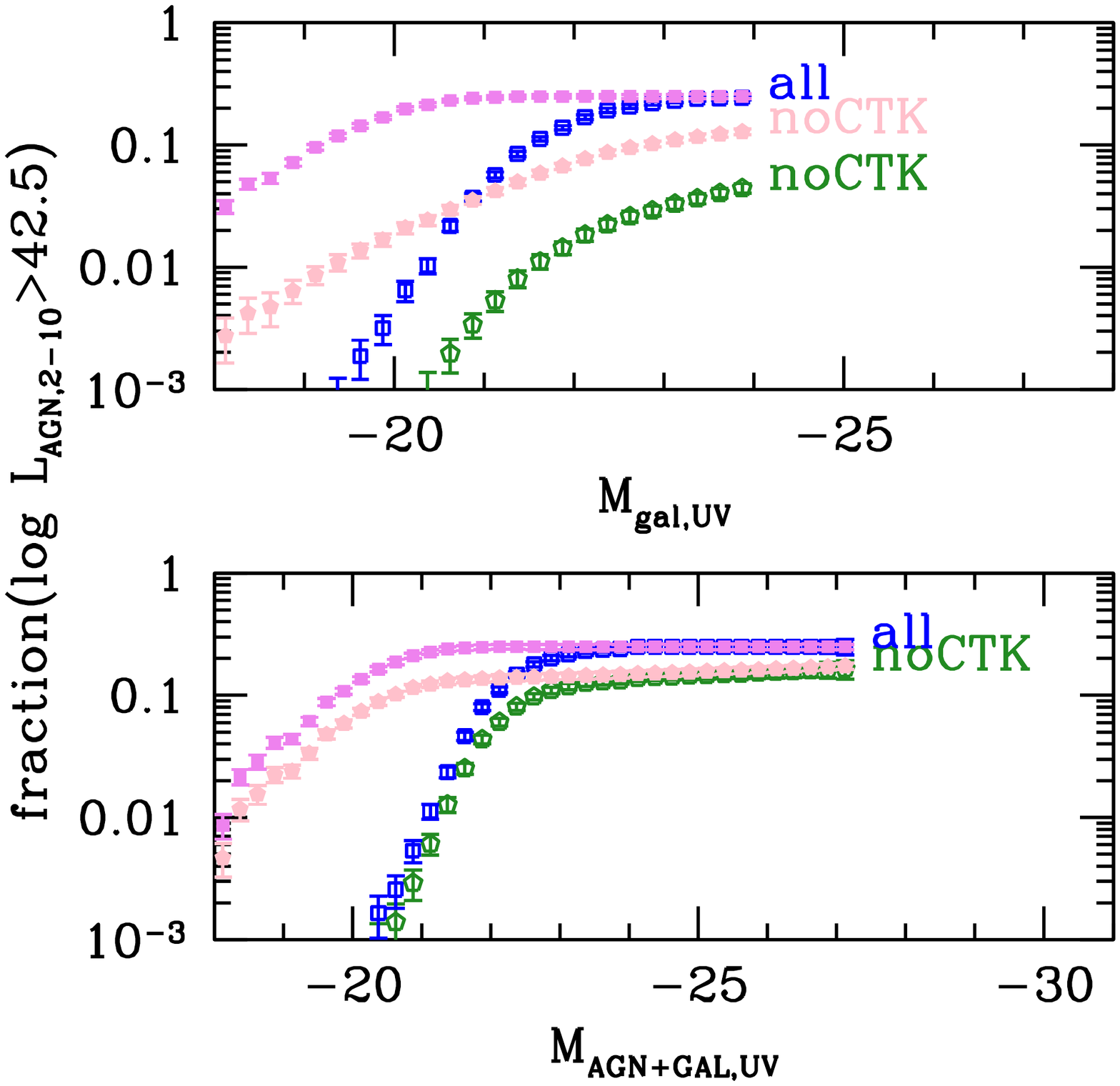}
\vspace{-1.cm}
\caption{\footnotesize Fraction of galaxies hosting an AGN with luminosity in the 2-10 keV band $>10^{42.5} \, {\rm erg\,s^{-1}}$  vs galaxy (top) and total (AGN+galaxy, bottom) UV magnitude. The empty (blue: all;  green: applying a correction for Compton Thick AGN) symbols refer to the reference model described in the body of the paper, the full (violet: all;  pink: applying a correction for Compton Thick AGNs) symbols to the super-maximal case.
\label{fig:frac_max_G}}
\end{figure}

To fit for the full range of the LF proposed by Giallongo et al. 2015, i.e., including the bright end, we modify the BH-galaxy relationship using the ``vanilla" scaling from \cite{2016ApJ...820L...6V}, i.e., $\mu=s-2.7$, and find $\bar{\lambda}=\log(0.75)$,  $\sigma_{\lambda}=0.4$,  $\sigma_{\mu}=0.2$. In this case, most galaxies with a UV magnitude $\sim -20$ should be AGN dominated, if they host an active BH (left panel of Fig.~\ref{fig:contamination_max_G}) and HMXBs would not be a significant contaminant even for BHs with $\mu=5$ (right panel of Fig.~\ref{fig:contamination_max_G}). Taking into account the assumed duty cycle of ${\cal D}=0.25$, 25\% of galaxies brighter than $-20$ are AGN-dominated, and in Fig.~\ref{fig:frac_max_G} we show the fraction of galaxies, as function of the galaxy (top) and total (bottom) UV magnitude where an X-ray detection is expected. In the reference model, most galaxies fainter than $-22$ are star-dominated, while in the super-maximal case a significant fraction, corresponding to all galaxies with an active BH, is AGN dominated down to a magnitude of $-20$.

\end{document}